\documentclass[12pt]{article}
\usepackage{amsmath,amssymb,bbold,hyperref,graphicx}
\usepackage{amsmath,xcolor}
\usepackage{mathtools}
\usepackage{physics}
\newcommand{\be}{\begin{equation}}
\newcommand{\ee}{\end{equation}}
\newcommand{\bea}{\begin{eqnarray}}
\newcommand{\eea}{\end{eqnarray}}
\newcommand{\beas}{\begin{eqnarray*}}
\newcommand{\eeas}{\end{eqnarray*}}

\begin{document}
\begin{titlepage}

\begin{center}

{\Large Entanglement and Chaos near Critical Point in Strongly Coupled Gauge Theory }

\vspace{12mm}

\renewcommand\thefootnote{\mbox{$\fnsymbol{footnote}$}}
Debanjan Karan \footnote{debanjan.karan@icts.res.in} and Sanjay Pant \footnote{2018phz0012@iitrpr.ac.in}


\vspace{6mm}

${}^*${\small \sl International Centre for Theoretical Sciences},
{\small \sl (ICTS-TIFR)},\\
{\small \sl Tata Institute of Fundamental Research, Shivakote, Hesaraghatta, Bengaluru 560089, India}

${}^{\dagger}$
{\small \sl Department of Physics},
{\small \sl Indian Institute of Technology Ropar},\\
{\small \sl Rupnagar, Punjab 140 001, India} \\

\end{center}

\vspace{12mm}

\noindent
     \hspace{7 cm}    \textbf{Abstract}\\

We perform a holographic study of the high and low temperature behaviours of logarithmic negativity (LN) and entanglement wedge cross section (EWCS) in a large $N$ strongly coupled thermal field theory with critical point, having a well defined gravity dual known as 1RC black hole. The bulk theory accommodates a dimensionless parameter $\xi$, proportional to the charge of the 1RC black hole. Holographically, $\xi \to 2$ limit ensures an existence of a critical point in the dual boundary theory. We show that the logarithmic negativity in low and high temperature limits enhances with increasing $\xi$. We analytically compute the EWCS in low and high temperature limits and find an agreement with the previously reported numerical results. We holographically explore the correlation between two identical copies of thermal field theory with critical point forming a thermofield double state (TFD) by computing the thermo mutual information (TMI). TMI shows an increasing behaviour with respect to the width of the boundary region. Furthermore, we study the chaotic behaviour of the field theory by analyzing a shock wave in the dual eternal 1 RC black hole and then estimate the degradation of TMI. The rate of such disruption of TMI slows down as the value of critical parameter $\xi$ takes higher values.

\end{titlepage}
\setcounter{footnote}{0}
\renewcommand\thefootnote{\mbox{\arabic{footnote}}}

\hrule
\tableofcontents
\bigskip
\hrule

\addtolength{\parskip}{8pt}
\newpage

\section{Introduction\label{sec:sec1}}

Recent development of quantum information theory truly advocates in favour of the quantum origin of gravity. A non-extremal black hole is a crucial object that provides a platform to explore the relation between gravity and the quantum information theory in a quantitative manner. Due to the non-extremality of the black hole, it becomes apparent that a suitable approach to recognize such relation is given by the study of the entanglement structure of mixed state in the bipartite formalism.  
A bipartite system is described by $\mathcal{H}_A \otimes\mathcal{H}_B$ where $\mathcal{H}_A$ and $\mathcal{H}_B$ are the Hilbert spaces of the individual subsystems $A$ and $B$ respectively. The entanglement entropy (EE) of the subsystem $A$ is defined as the
von Neumann entropy, 
\begin{equation}\label{eq11}
    \mathcal{S}_A = - \text{Tr}( \rho_A \log \rho_A),
\end{equation}
where, $\rho_A= \text{Tr}_B( \rho_{AB})$ is the reduced density matrix of $A$ and $\rho_{AB}$ is the total density matrix defined in the bipartite Hilbert space $\mathcal{H}_A \otimes\mathcal{H}_B$ \cite{Chuang}. EE is not a reliable measure of quantum correlation when $A$ and $B$ are non-complementary to each other. In such a case, there exists a positive semi definite measure known as mutual information (MI), $   I(A:B)= \mathcal{S}_A+ \mathcal{S}_B-\mathcal{S}_{A\cup B}$ that correctly characterizes both classical and quantum correlation in a  mixed bipartite entangled state \cite{Winter}.

However, $I(A: B) = 0$ implies that the non-complementary subsystems may or may not be entangled and in that case, additional entanglement measures or criteria are required. Apart from MI, other measures such as entanglement of purification (EoP) and logarithmic negativity (LN) are widely used to diagnose the entanglement for a mixed state \cite{Terhal,Peres:1996dw}.

According to the concept of purification, starting with a mixed density matrix operator $\rho_{AB}$ defined in 
 a bipartite Hilbert space $\mathcal{H}_A \otimes \mathcal{H}_B $ of two noncomplementary systems $A$ and $B$, one can prepare a pure state $\ket{\psi}$ in an extended Hilbert space represented as $\mathcal{H}_A \otimes \mathcal{H}_B\otimes \mathcal{H}_{A'}\otimes \mathcal{H}_{B'} $.  It is important to note that there exists an infinite arrays of purification pathways for $\ket{\psi}$, all satisfying the condition that $\rho_{AB}=Tr_{A'B'}\ket{\psi}\bra{\psi}$. For a given bipartite mixed state $\rho_{AB}$, we define the EoP, denoted as $ E_p(A:B)$, as the minimum of EE among all feasible purification \cite{Terhal},
\begin{equation}
    E_p(A:B) = min_{\rho_{AB} = Tr_{A'B'} |\Psi\rangle \langle \Psi|} \{\mathcal{S}(\rho_{AA'}) \}
\end{equation}

In practice, the difficulties with the choice of purification pathways can be bypassed by 
introducing logarithmic negativity (LN) as a measure of the upper bound on the distillable entanglement in a mixed state \cite{Vidal:2002zz}
\begin{equation}
    \mathcal{E}=\log ||\rho_{AB}^T||
\end{equation}
where $\rho^T_{AB}$ is partial transpose of the mixed density matrix $\rho_{AB}$ with respect to the subsystem $B$ and $||\rho_{AB}^T||$ is the trace norm. Trace norm is directly related to the entanglement negativity via $ N= \frac{||\rho_{AB}^T||-1}{2}$ \cite{Vidal:2002zz}. Further, the convexity issue of LN was addressed in \cite{Plenio:2005cwa}. Unlike the MI, LN is defined in such a way that it can extract the quantum correlation alone present in a bipartite mixed state. 

Field theoretic analysis for EoP hardly exists due to the difficulties in the implementation of the minimization procedure except for some numerical results found in free lattice field theory \cite{Bhattacharyya:2018sbw}.
On the otherhand, LN has been computed in CFT$_2$ by employing a version of the usual replica technique that involves a specific four-point function of the twist and anti-twist fields. \cite{Calabrese:2012nk, Calabrese:2012ew, Calabrese:2014yza, Calabrese:2013mi, Chung:2014ed, Wen:2015qwa, Caste:2013nt, Coser:2014gsa}.
 
The direct study of entanglement measures for strongly coupled field theory is still an open question due to the non-perturbative nature of the theory. Nevertheless, one can still study the entangled structure of such strongly coupled systems by following appropriate prescriptions provided by the  holographic duality.

A concrete example of holographic duality is the AdS/CFT correspondence that suggests the information of a conformal field theory (CFT) living on the boundary of Anti-de Sitter (AdS) space is encoded in the bulk gravitational theory of AdS \cite{Maldacena:1997re, Witten:1998qj}. 
 Although general proof of the conjecture is yet to achieve, it passes numerous consistency tests in diverse fields. The Ryu-Takayanagi formula is a crucial example in favour of the AdS/CFT correspondence and it provides a holographic prescription for computing the entanglement entropy in the boundary CFT, known as holographic entanglement entropy (HEE)\cite{Ryu:2006bv, Ryu:2006ef}.  
It states that the entanglement entropy of a certain region A in the CFT is given by the area of a minimal surface (called the Ryu-Takayanagi surface) in the bulk AdS spacetime that is homologous to the boundary region.
\begin{equation}\label{eq12}
    \mathcal{S}_A = \frac{\mathcal{A}(\gamma_A)}{4G_N^{d+1}}
\end{equation}
where $\gamma_A$ is a co-dimension two surface with the area $\mathcal{A}(\gamma_A)$ such that
${\partial}{\gamma_A} = \partial A$ and $G_N^{d+1}$ is the $d+1$ dimensional Newton's constant. Later Hubeny, Rangamani, and Takayanagi (HRT) extended this idea for time dependent spacetime \cite{Hubeny:2007xt}.
The study of entanglement entropy in the context of AdS/CFT has provided valuable insights into quantum phase transitions and critical phenomena in both the boundary CFT and the bulk gravity theory \cite{Fradkin:2006mb, Osborne:2002zz, Baggioli:2023ynu}.

 Quite naturally constructing a holographic prescription of computing the entanglement structure in a bipartite mixed entangled state is needed.
In the context of $AdS_3/CFT_2$, the authors of \cite{Chaturvedi:2016rcn} propose a holographic conjecture to compute the LN of such boundary CFTs that exactly reproduces the CFT$_2$ results in large central charge limit \cite{Calabrese:2014yza}. See \cite{Jain:2017aqk,Afrasiar:2021hld} for further generalization of this proposal.
Similarly, a viable holographic measure dual to the EoP was proposed in \cite{Takayanagi:2017knl, Nguyen:2017yqw}. In  \cite{Takayanagi:2017knl} this holographic dual named as the entanglement wedge cross section (EWCS). It is important to note that the EWCS is also related to the mutual information and always greater the equal to the half of the MI.

 Entanglement and quantum chaos are two distinct concepts, but they are interconnected in various ways, especially when we study them from the perspective of a mixed state. In particular, we start with two identical copies of causally disconnected but non-locally correlated thermal field theories, use the notion of purification and construct a TFD state which has the holographic dual, a two sided eternal black hole \cite{Maldacena:2001kr}. An elegant way to quantify such non-local correlation present in the TFD state is to study a variant of MI, known as theormo-mutual information (TMI) \cite{Morrison:2012iz}. TMI counts the total non-local correlation between two causally disconnected thermal field theories. On the other hand, in \cite{Shenker:2013pqa}, the authors show that in a chaotic system the entanglement between two causally disconnected parts of a TFD state can be disrupted by an early perturbation which grows exponentially in time. The generalization of similar study for a strongly coupled field theory is executable via various  holographic prescriptions such as shockwave analysis and pole skipping method \cite{Shenker:2013pqa, Blake:2018leo, Grozdanov:2018kkt, Jahnke:2017iwi, Sil:2020jhr, Mahish:2022xjz,Chakrabortty:2022kvq}.

In this work we mainly focus on the analysis of  entanglement measure of mixed state and its connection with quantum chaos from the  perspective of  the four dimensional strongly coupled $\mathcal{N} = 4$ super Yang-Mills theory charged under a $U(1)$ subgroup of its $SU(4)$ R-symmetry at finite temperature and  chemical potential via the holographic computation in the dual five dimensional 1RC black hole background  \cite{Gubser:1998jb, Cvetic:1998jb}.
The analysis of thermodynamic stability in the strongly coupled field theory we consider here can be carried out in a one dimensional phase space characterized by a dimensionless ratio $\frac{\mu}{T}$ \cite{Ebrahim:2020qif}. Such an analysis ensures the existence of a critical point that corresponds to $\xi \to 2$ where $\xi$ is a dimensionless parameter in the dual five dimensional 1RC black hole background \cite{Ebrahim:2020qif}. It is very interesting to probe such critical point by the measure of mixed state entanglement as an order parameter of phase transition. Motivated by this fact, the low and high temperature limits of HEE and HMI near the critical point have been explored in 1RC black hole background\cite{Ebrahim:2020qif}. The author show that at and near the critical point the leading behavior of mutual information yields a set of critical exponents. Moreover, in \cite{Amrahi:2020jqg}, a numerical investigation of the EWCS holographically reveals that the EoP in the dual field theory at finite temperature ($T$) and chemical potential ($\mu$) 
behaves as a monotonic function of $\frac{\mu}{T}$
whereas the EoP behaves drastically different in the presence of a critical point. The analysis of the holographic butterfly effect is carried out within the background of a 1RC black hole. In this context, the dynamical exponent is determined through an expansion of the butterfly velocity in the vicinity of the critical point, as described in \cite{Amrahi:2023xso}. See \cite { Kraus:1998hv, Cvetic:1999ne, Cvetic:1999rb, Ebrahim:2018uky, Ebrahim:2017gvk}, for more holographic applications in this background. 

In the present analysis, we aim to improve the understanding of classical and quantum correlations of mixed state near the critical point of strongly coupled $\mathcal{N} = 4$ super Yang-Mills theory at finite temperature and chemical potential, via performing holographic computation of various relevant measures such as LN, EoP and TMI in the dual five dimensional 1RC black hole background. In our analysis we find that, for adjacent as well as bipartite configurations, the HLN increases with $\xi$ parameter for both high and low values of temperature.  Crucially, for each set up, HLN remain finite in critical limit $\xi \to 2$, however the slope of HLN exhibits a power law divergence at the critical point. We also observe that EoP  increases with respect to the parameter $\xi$ and remains finite in the critical limit. We further show that TMI between two entangled subsystems forming a TFD state increases with their individual sizes. At a fixed size of the subsystem, TMI rises with increasing $\xi$. In order to carry through our investigation exploring the chaotic dynamics of strongly coupled field theories featuring critical point, we introduce a time-dependent perturbation in the asymptotic past. This perturbation, when realized within the holographic framework, takes the form of an exponentially growing energy pulse and ultimately manifests itself as a shock wave. We notice a disruption of the holographic TMI in the presence of shockwave, and our analysis indicates that as $\xi$ parameter takes higher values, the chaotic behavior of the  system gets reduced.
 
 This paper is organized as follows; in section \ref{sec:sec2} we discuss about the holographic dual of the strongly coupled field theory with critical point. In section \ref{sec:sec3} we review the HEE. Sections \ref{sec:sec4}, \ref{sec:sec5} and \ref{sec:sec6} are devoted to the holographic logarithmic negativity (HLN) for two subsystems in different configurations. In section \ref{sec:sec7} we give the analytic form of EWCS in low and high thermal limits and in section \ref{sec:sec8} we present a computation of thermo-mutual information. Finally, in section \ref{sec:sec9} we summarize our result and conclude.

\section{Background}\label{sec:sec2}

In this section we briefly review some important features of five-dimensional 1RC black hole \cite{Gubser:1998jb,Cvetic:1998jb,Kraus:1998hv, Cvetic:1999ne,Cvetic:1999rb}. 
Consider the following five dimensional  Einstein-Maxwell-dilaton action
 \begin{equation}\label{eq1}
    \mathcal{S}_{\text {EMD}}
=
    \frac{1}{16 \pi G^{(5)}_{N}} \int d^5 x \sqrt{-g}\left[\mathcal{R}-\frac{f(\phi)}{4}F_{\mu \nu}F^{\mu \nu}-\frac{1}{2}\partial_{\mu}\phi \partial^{\mu}\phi-V(\phi)\right],
 \end{equation}
 where $A_\mu$ is the gauge field and $\phi$ is a scalar field. We denote the dilaton potential as $V(\phi)$ and the coupling between the gauge field and the dilaton is characterized by the coupling function $f(\phi)$. The functions $ f(\phi)$ and $V(\phi)$ have the following form
 
\begin{equation}\label{eq2}
    f(\phi)=e^{-\sqrt{\frac{4}{3}}\phi}, ~~~
    V(\phi)=-\frac{1}{R^2}\left( 8e^{\frac{\phi}{\sqrt{6}}} + 4 e^{-\sqrt{\frac{2}{3}}\phi} \right)
\end{equation}
where $R$ is the $AdS$ radius. The solution to the equations of motion of the EDM action corresponds to the 1RC black hole background described by
 \begin{equation}\label{eq3a}
   d s^{2}= e^{2A(z)} \left(-h(z) d t^{2}+d \vec{x}^{2}_{(3)} \right) + \frac{e^{2B(z)}}{h(z)} \frac{R^4}{z^4} dz^2
 \end{equation}
where, 
\begin{equation}\label{eq4}
\begin{split}
A(z)&= \ln\left( \frac{R}{z} \right) + \frac{1}{6} \ln \left( 1 + \frac{Q^2 z^2}{R^4} \right) \\
B(z)&= -\ln\left( \frac{R}{z} \right) - \frac{1}{3} \ln \left( 1 + \frac{Q^2 z^2}{R^4} \right) \\
h(z)&= 1 - \frac{M^2z^4}{R^6 \left(  1 + \frac{Q^2 z^2}{R^4}  \right)},
\end{split}
\end{equation} and the corresponding scalar field $\phi(z)$ and the electric potential $\Phi(z)$ that attributes to the temporal component of the gauge field are described by
\begin{equation}
\begin{split}
\label{eq3b}
\phi(z)&= -\sqrt{\frac{2}{3}} \ln \left( 1 + \frac{Q^2 z^2}{R^4} \right),  \\
\Phi (z)&= \frac{MQ {z_h}^2}{R^4 \left( 1 + \frac{Q^2 {z_h}^2}{R^4} \right)} - \frac{MQ {z}^2}{R^4 \left( 1 + \frac{Q^2 {z}^2}{R^4} \right)}.
\end{split}
\end{equation} 
Here we are working with Poincare's coordinate system where the boundary is situated at $z= 0$ and the horizon is given by $h(z)=0$. Note that, the electric potential $\Phi(z) $ is chosen in such a way that it is regular on the boundary \cite{DeWolfe:2010he, DeWolfe:2011ts}  and vanishes on the horizon. The parameters $M$ and $Q$ are related to the mass and charge  of the black hole respectively. Moreover, the blackening function $h(z)$ takes the following form,
\begin{equation}\label{eq5}
    h(z)= 1 - {\left( \frac{z}{z_h}\right) }^4 \left( \frac{1 + {(\frac{Qz_h}{R^2})}^2}{1+ {(\frac{Qz}{R^2})}^2} \right) =1 - {\left( \frac{z}{z_h}\right) }^4 \left( \frac{1 + \xi}{1+ \xi(\frac{z}{z_h})^2} \right) 
\end{equation}
where, $\xi \equiv Q^2 z_h^2/R^4$.

The Hawking temperature is given by,

\begin{equation}\label{eq8}
    T=\frac{1}{2\pi z_h} \left( \frac{ 2 + {\left(\frac{Qz_h}{R^2}\right)}^2}{\sqrt {1 + {\left(\frac{Qz_h}{R^2}\right)}^2}}\right)
\end{equation}
and the chemical potential is, 
\begin{equation}\label{eq9}
    \mu = \frac{1}{R} \lim_{z\to  0} \Phi(z) = \frac{Q}{R^2 \sqrt{1 + {\left(\frac{Qz_h}{R^2}\right)}^2}}.
\end{equation}
For convenience, we rewrite the temperature, in terms of a dimensionless quantity $\xi$ and the effective temperature $\hat{T}$ as
\begin{equation}\label{eq10}
    T=\hat{T} \left( \frac{1 + \frac{\xi}{2}}{\sqrt{1+\xi}} \right),~~~\hat{T} \equiv \frac{1}{\pi z_h}
\end{equation}
 
\begin{figure}
\centering
\includegraphics[width=.50\linewidth]{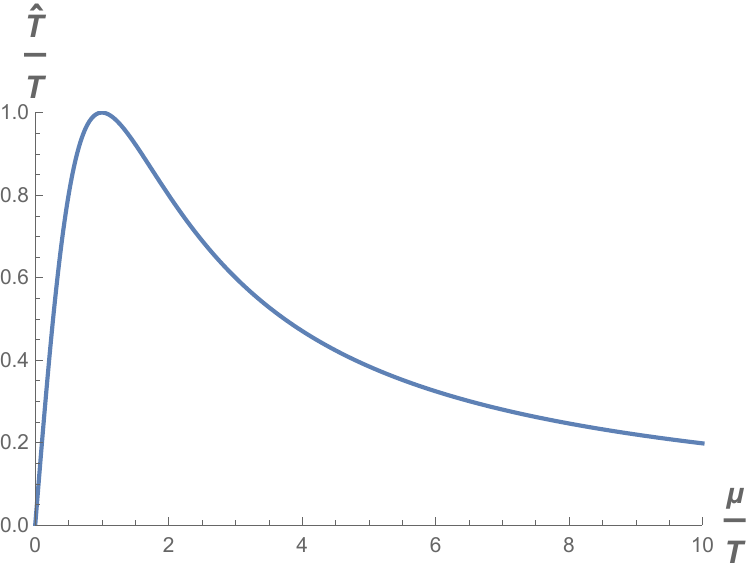}
\caption{ $\hat{T}/T$ vs $\mu/T$ for $\frac{TR^2}{Q}=1$.}
\label{Tmu}
\end{figure}

 Using equation \eqref{eq9} and \eqref{eq10} we can write the effective temperature in terms of the $\mu$, $Q$ and $T$ as,
\begin{equation}\label{Tfmu}
    \frac{\hat{T}}{T}=\frac{2 T R^2}{Q}\frac{\left(\frac{\mu}{T}\right)}{\left[1+{\left(\frac{TR^2}{Q}\right)}^2\left(\frac{\mu}{T}\right)^2\right]}
\end{equation}
For the small values of $\frac{\mu}{T}$ i.e $\frac{\mu}{T}\ll 1$ we have,
\begin{equation}\label{sTfmu}
    \frac{\hat{T}}{T}=\frac{2 T R^2}{Q}\left(\frac{\mu}{T}\right)-\mathcal{O}{\left(\frac{\mu}{T}\right)}^3
\end{equation}
and for the large values $\frac{\mu}{T}\gg 1$, equation \ref{Tfmu} becomes
\begin{equation}\label{hTfmu}
    \frac{\hat{T}}{T}=\frac{2 Q}{T R^2}\frac{1}{\left(\frac{\mu}{T}\right)}
\end{equation}
Figure \ref{Tmu} demonstrates the behaviour of $\hat{T}/T$ with respect to ${\mu}/{T}$. It is evident from the figure \ref{Tmu} that in the low $\mu/T$ the quantity $\hat{T}/T$ increases linearly which is consistence with equation \eqref{sTfmu}. Also for high values of $\mu/T$ the ratio $\hat{T}/T$ shows a decrement which is expected from the equation \eqref{hTfmu}. 

The entropy density $(s)$ and $U(1)$ charge density $(\rho)$ are given by,
 \begin{equation}
     s=\frac{R^3}{4G_N z_h^3}\sqrt{1+\xi},~~~~~~~~~ \rho=\frac{QR}{8\pi G_N z_h^2}\sqrt{1+\xi}
 \end{equation}
The specific heat at constant chemical potential, $C_\mu$ and the second order R-charge susceptibility at constant temperature $\chi_2$ ca be evaluated as,
 \begin{equation}
     -\biggl(\frac{\partial^2f}{\partial T^2}\biggr)_{\mu}=\biggl(\frac{\partial s}{\partial T}\biggr)_{\mu}\equiv \frac{C_{\mu}}{T},~~~~ -\biggl(\frac{\partial^2f}{\partial \mu^2}\biggr)_{T}=\biggl(\frac{\partial \rho}{\partial T}\biggr)_{T}\equiv\chi_2. 
 \end{equation}
 Most importantly, at $\frac{\mu}{T}=\frac{\pi}{\sqrt{2}}$ both of the above quantities diverge. The phase structure of the field theory dual to the 1RCBH background exhibits a second order phase transition and the critical point is characterized by $\frac{\mu_c}{T_c}=\frac{\pi}{\sqrt{2}}$ which is same as $\xi=\frac{Q^2z_h^2}{R^4}=2$
\cite{Ebrahim:2020qif}. It is convenient 
 for future discussions to write $\xi$ in terms of the temperature and chemical potential as,
 \begin{equation}\label{xil}
 \xi=\frac{2(1-\sqrt{1-\lambda^2})^2}{\lambda^2}
 \end{equation}
  where, $\lambda \equiv \frac{(\mu / T)}{(\pi / \sqrt{2})}$ and the critical point is defined by $\lambda \to 1$.

\section{Holographic Entanglement Entropy (HEE)}\label{sec:sec3}

In this section, we summarize the computation of HEE at low and high values of temperature as presented in \cite{Ebrahim:2020qif}. Subsequently, we employ the method outlined in \cite{Jain:2017xsu, KumarBasak:2020viv} to compute the HLN for the 1RCBH background. To elaborate, we consider a boundary subsystem in the shape of a rectangular strip denoted as $A$ with a width $l$ along the $x$ direction and the length $L$ in all the transverse directions $x^j$.  The rectangular strip $A$ is precisely defined as follows,

\begin{equation}\label{eq13}
    x \equiv x \in \Big[-\frac{l}{2},\frac{l}{2}\Big], \hspace{5mm} x^{j} \in \Big[-\frac{L}{2},\frac{L}{2}\Big], ~~~ j= 2, 3 
\end{equation}
where $L$ is considered very  large as compared to $l$. Determining the HEE of subsystem $A$ requires us to calculate the smallest surface area of the co-dimension two hyper-surface denoted as ${\gamma}_A$. The area functional of ${\gamma}_A$ is as follows:

\begin{equation}\label{eq14}
    \mathcal{A}({\gamma}_A)=\int d^{3}x~\sqrt{\text{det}(g_{mn})}
\end{equation}
where $g_{mn}$ is the induced metric of $\gamma_A$. By using the explicit form of induced metric, we express the area functional as,
\begin{equation}\label{eq15}
 \mathcal{A}=2L^2 \int dz~e^{3A(z)} \sqrt{{x'(z)}^2 + \frac{R^4}{z^4 h(z)} e^{2(B(z)-A(z))}}
\end{equation}
  One can find the conserved quantity corresponds to $x$ using the Lagrangian in the area functional and obtain the following equation by imposing $\frac{1}{x'(z_t)}=0$ for $z\to z_t$,
 \begin{equation}\label{eq16}
     x'(z) = \frac{R^2}{z^2} \frac{e^{3A(z_t)} e^{B(z)-A(z)}}{\sqrt{h(z)}\sqrt{e^{6A(z)} - e^{6A(z_t)}}}
 \end{equation}
where  $z_t$ is the turning point of the surface ($\gamma_A$). Using $x'(z)$, the area functional \eqref{eq15} now becomes
\begin{equation} \label{eq17}  \mathcal{A}=2L^{2}R^{2}\int_{0}^{z_t}dz~\frac{e^{B(z)+2A(z)}}{z^{2}\sqrt{h(z)}}\sqrt{\frac{e^{6A(z)}}{e^{6A(z)}-e^{6 A(z_t)}}}
\end{equation}
Finally substituting \eqref{eq4} into \eqref{eq17}, the holographic entanglement entropy (HEE) takes the form,
 \begin{equation}\label{eq18}
   \mathcal{S}=\frac{L^{2}R^{2}}{2G_N^5}\int_{0}^{z_t}dz~\frac{e^{B(z)+2A(z)}}{z^{2}\sqrt{h(z)}}\sqrt{\frac{e^{6A(z)}}{e^{6A(z)}-e^{6 A(z_t)}}}
 \end{equation}
 From \eqref{eq16} the boundary parameter $l$ and the bulk parameter $z_t$ are related via, 
\begin{equation}\label{lzt}
     \frac{l}{2} = \int_{0}^{z_t}dz~\frac{R^2}{z^2} \frac{e^{3A(z_t)} e^{B(z)-A(z)}}{\sqrt{h(z)}\sqrt{e^{6A(z)} - e^{6A(z_t)}}}
 \end{equation}
 To express the HEE in terms of boundary parameter we have to replace the $z_t$ in \eqref{eq18} in terms of $l$. Finding a solution for the integral \eqref{lzt} and expressing $z_t$ in relation to $l$ poses a significant challenge. Nevertheless, at low and high values of temperature, accomplishing this task becomes feasible. 
 
To express the area functional in terms of black hole parameters we substitute \eqref{eq4} into \eqref{eq17} and obtain,
\begin{equation}\label{eq20}
     \mathcal{A} = 2L^2 R^3 \int_0^{z_t} dz ~ \frac{{z_t}^3}{{z}^6}\sqrt{\frac{1+\xi {(\frac{z}{z_h})}^2}{1+\xi {(\frac{z_t}{z_h})}^2}} 
     {\left[ 1- {\left(\frac{z}{z_h}\right)}^4\left( \frac{1 + \xi}{1+\xi {(\frac{z}{z_h})}^2} \right)\right]}^{-\frac{1}{2}}
     {\left[ {\left( \frac{z_t}{z} \right)}^6 \left( \frac{1+\xi {(\frac{z}{z_h})}^2}{1+\xi {(\frac{z_t}{z_h})}^2} \right) - 1 \right]}^{-\frac{1}{2}}
 \end{equation}

 In a similar way equation \eqref{lzt} can be expressed as,
\begin{equation}\label{eq21}
    \frac{l}{2}= \int_0^{z_t} dz~ {\left[ 1 + \xi {\left( \frac{z}{z_h}\right)}^2 \right]}^{-\frac{1}{2}}
    {\left[ 1- {\left(\frac{z}{z_h}\right)}^4\left( \frac{1 + \xi}{1+\xi {(\frac{z}{z_h})}^2} \right)\right]}^{-\frac{1}{2}}
     {\left[ {\left( \frac{z_t}{z} \right)}^6 \left( \frac{1+\xi {(\frac{z}{z_h})}^2}{1+\xi {(\frac{z_t}{z_h})}^2} \right) - 1 \right]}^{-\frac{1}{2}}
\end{equation}
\begin{figure}
\centering
\includegraphics[width=.50\linewidth]{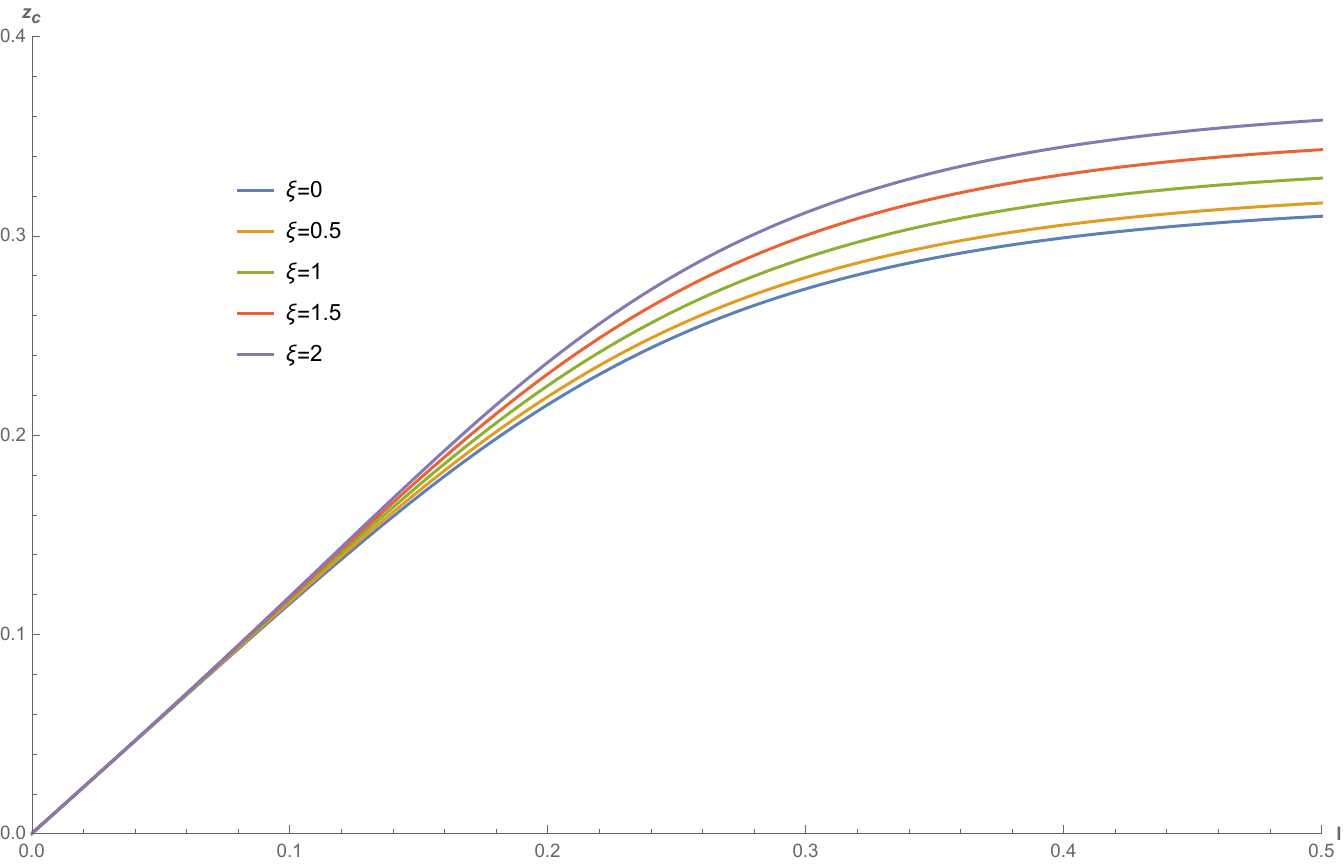}
\caption{Turning point $z_t$ of RT surface with respect to width $l$.}
\label{zcl}
\end{figure}
To evaluate the integrals \eqref{eq20} and \eqref{eq21} we employ the following series expansion formulae,
\begin{equation}\label{eq22}
\begin{split}
    &{(x+y)}^{-n} = \sum_{k=0}^\infty{(-1)}^k \frac{\Gamma(n+k)}{\Gamma(k+1)\Gamma(n)} x^{-n-k}y^k;~~\text{given}~\vert y\vert<\vert x\vert \\
    &{(x+y+z)}^{-n} = \sum_{k=0}^\infty\sum_{j=0}^k  \frac{\Gamma(n+k)}{\Gamma(k+1)\Gamma(n)} \frac{{(-1)}^k\Gamma(k+1)}{\Gamma(j+1)\Gamma(k-j+1)}x^{-n-k} y^{k-j}z^j,~~\text{given}~\vert y+z \vert<\vert x\vert
\end{split}
\end{equation}
Using equation \eqref{eq22} we can write the following form of the area integral
\begin{equation}\label{eq23}
\begin{split}
\mathcal{A} = \frac{2L^2R^3}{\pi} \sum_{k=0}^{\infty} \sum_{n=0}^{k} \sum_{m=0}^{\infty} \sum_{j=0}^{\infty} \frac{{(-1)}^{k+n}\Gamma(k + \frac{1}{2}) \Gamma (j+m+ \frac{1}{2}) }{ \Gamma (n+1) \Gamma (k-n+1) \Gamma (j+1) \Gamma (m+1)} {\xi}^{k-n+m} {(1+\xi)}^n {\left( \frac{z_t}{z_h}\right)}^{2m}  \\
\times {\left[ 1 + \xi { \left( \frac{z_t}{z_h} \right)}^2 \right]}^{-m-\frac{1}{2}} \int_0^{{\color{red}z_c}} dz~ \left[ 1 + \xi {\left(\frac{z}{z_h}\right)}^2\right] \left[ 1- {\left(\frac{z}{z_t}\right)}^2\right] z^{-3} {\left( \frac{z}{z_t}\right)}^{6j} {\left( \frac{z}{z_h} \right)}^{2(k+n)} 
\end{split}
\end{equation}

The region bounded by the extremal surface exhibits divergence, primarily due to its behavior near the boundary. Upon closer examination, it becomes evident that when the condition $k + n + 3j > 1$ is met, the final integral (and consequently, the enclosed area) remains finite. Consequently, we must isolate and sum the terms corresponding to $(k = n = j = 0)$ and $(k = 1, n = j = 0)$ over the variable $m$ to determine the portion of the region containing the divergent component. By carrying out this procedure, one can derive the subsequent outcome.
\begin{equation}\label{eq24}
    {\mathcal{A}}_0 \equiv L^2R^2 \Bigg{\{} \frac{1}{{\epsilon}^2} + \frac{3\xi}{2{z_h}^2} - \frac{1}{{z_t}^2} {\left[  1 
 + \xi {\left( \frac{z_t}{z_h} \right)}^2 \right]}^{\frac{3}{2}}  \Bigg{\}}
\end{equation}

Note that, here we use a cutoff surface located at $z=\epsilon$ within the bulk geometry and the $\epsilon$ parameter holographically corresponds to the ultraviolet (UV) regularization parameter of the dual field theory. It becomes evident that the divergent term in equation \eqref{eq24} shows behavior akin to an area law. Naturally, the same characteristic is also shared by the associated holographic entanglement entropy. In the context of a $d$-dimensional boundary field theory, when the leading divergent term in the UV limit, as $\epsilon$ tends towards zero, adheres to an area law, this outcome is entirely anticipated. For simplicity, we  focus on the finite component of the area, achieved by subtracting the $1/\epsilon^2$ term. This can be expressed in the following manner:
\begin{multline}\label{eq25}
     {\mathcal{A}}_{\text{finite}} = \frac{L^2 R^3}{{z_t}^2}\Bigg{\{} \frac{3\xi}{2} {\left( \frac{z_t}{z_h}\right)}^2 - {\left[ 1 + \xi {\left( \frac{z_t}{z_h}\right)}^2\right]}^{\frac{3}{2}} + \frac{1+\xi}{3\xi}{\left( \frac{z_t}{z_h}\right)}^2 \left[ {\left( 1 + \xi {\left( \frac{z_t}{z_h}\right)}^2 \right)}^{\frac{3}{2}} - 1 \right]\Bigg{\}} \\
     + \frac{L^2 R^3}{{z_t}^2} \Bigg{\{} \sum_{k=2}^{\infty} \sum_{n=0}^k \sum_{m=0}^{\infty} {\Lambda}_{knm} \frac{ \Gamma(m+\frac{1}{2})\Gamma(k+n-1)}{\Gamma(k+n+m+1)} {\left( \frac{z_t}{z_h} \right)}^{2(k+n+m)} \\ \times \left[ (m+1) + (k+n-1) \left( 1 + \xi{\left( \frac{z_t}{z_h}\right)}^2 \right) \right]  \Bigg{\}} \\
     ~~~~~~~~~~~+ \frac{L^2 R^3}{{z_t}^2} \Bigg{\{} \sum_{k=0}^{\infty} \sum_{n=0}^k \sum_{m=0}^{\infty} \sum_{j=1}^{\infty} {\Lambda}_{knm} \frac{ \Gamma(m+j+\frac{1}{2})\Gamma(k+n+3j-1)}{\Gamma(j+1)\Gamma(k+n+m+3j+1)} {\left( \frac{z_t}{z_h} \right)}^{2(k+n+m)} \\  \times \left[ (m+1) + (k+n+3j-1) \left( 1 + \xi{\left( \frac{z_t}{z_h}\right)}^2 \right) \right]  \Bigg{\}}
\end{multline}
where ${\Lambda}_{knm}$ is given by the following relation
\begin{equation}\label{eq26}
    {\Lambda}_{knm} \equiv \frac{ {(-1)}^{k+n} \Gamma(k+\frac{1}{2})}{\pi \Gamma (n+1) \Gamma (k-n+1)} {\xi}^{k-n+m} {\left( 1+\xi \right)}^n {\left[ 1 + \xi {\left( \frac{z_t}{z_h}\right)}^2  \right]}^{-m-\frac{1}{2}}
\end{equation}
Hence, by incorporating the UV-divergence-dependent term into \eqref{eq26}, we can derive the overall surface area of the extremal surface associated with a rectangular strip having a width of $l$ on the boundary. Similarly, by following this procedure, we can determine the subsystem's width as a function of the turning point. Further, through the utilization of multinomial expansions and the solution of the integral, outlined in equation \eqref{eq21}, the following relation is established,

\begin{equation}\label{eq27}
    \frac{l}{2} = z_t \sum_{k=0}^{\infty} \sum_{n=0}^{k} \sum_{m=0}^{\infty} \sum_{j=0}^{\infty} G_{knmj} F_{knmj} {\left( \frac{z_t}{z_h} \right)}^{2(k+n+m)} 
\end{equation}
where the constants $G_{knmj}$ and $F_{knmj}$ are defined as,
\begin{equation}\label{eq28}
\begin{split}
    G_{knmj} &\equiv \frac{\Gamma(k + \frac{1}{2}) \Gamma (j+m+ \frac{1}{2}) \Gamma(2+3j+k+n)}{2\pi \Gamma (n+1) \Gamma (k-n+1) \Gamma (j+1) \Gamma (3+3j+k+n+m)} \\
    F_{knmj} &\equiv {(-1)}^{k+n} {\xi}^{k-n+m} {(1+\xi)}^n {\left[ 1 + \xi { \left( \frac{z_t}{z_h} \right)}^2 \right]}^{-m}.
\end{split}
\end{equation}

In order for the multinomial expansions to hold true for negative exponents, the following inequalities must be satisfied for values of $z_t$ spanning from the boundary to the horizon and for the entire interval $\xi \in [0,2]$,
\begin{equation}\label{eq29}
    \frac{\xi {\left( \frac{z_t}{z_h} \right)}^2}{1 + \xi {\left( \frac{z_t}{z_h} \right) }^2} \left( 1 -\frac{z^2}{{z_t}^2} \right)< 1,~~~~~~\text{and}~~~~~~ \xi {\left( \frac{z}{z_h} \right)}^2 - (1+\xi){\left( \frac{z}{z_h} \right)}^4 < 1
\end{equation}

By examining equation \eqref{eq25}, we  note that the extremal surface area is characterized by two dimensionless parameters: $\xi$ and the ratio $z_t/z_h$. 
To express the final results only in terms boundary parameters, one can invert the relation \eqref{eq27} and express $z_t$ in terms of $l$ in the low and high effective temperature limits.
From the perspective of field theory, the low and the high temperature limits are defined as : $\hat{T}l \ll 1$ and $\hat{T}l \gg 1$, respectively, where $\hat{T}$ is defined in equation \eqref{eq10}.
In order to holographically probe the same limits we consider a dimensionless ratio between the location of the turning point  to the position of the horizon, denoted as $z_t/z_h$ and  focus on two distinct regimes of the parameter, e.g.,  $z_t/z_h \ll 1$ and $z_t/z_h \sim 1$. It is important to note that $z_t/z_h \ll 1$  implies that the extremal surface is situated close to the boundary at $z=0$, while the $z_t/z_h \sim 1$  indicates that it approaches to the horizon.  Consequently, one can associate the case of $z_t/z_h \ll 1$ with the low-temperature limit, corresponding to the ground state of the CFT, and the case of $z_t/z_h \sim 1$ with the high-temperature limit, where the thermal excitation becomes significant. 
 For both low and high temperatures, one can further impose the limit $\xi \to 2$ to probe the properties of HEE near the critical point of the theory. From \eqref{eq25} it can be seen that the area functional is finite in the limit $\xi \to 2$.
Moreover, as pointed out in the Fig.\ref{zcl}, when the parameter $l$ approaches zero, the turning points of the RT surfaces associated with different values of $\xi$ are hardly distinguished. However, beyond a certain threshold value of $l,$ such distinctions of turning points are apparent. It is important to note that, for a fixed $\xi$, the turning point initially emerges from the origin and gradually increases as $l$ increases. This behavior indicates that as the width of the boundary region expands the RT surfaces extend deeper into the bulk of the system. The fact that $z_t$ saturates for higher values of $l$ signifies that the RT surface, associated with a boundary region of width $l$, becomes nearly parallel to the horizon.

\section{Holographic Logarithmic Negativity for two adjacent subsystems}\label{sec:sec4}

\begin{figure}
\centering
 \includegraphics[width=.60\linewidth]{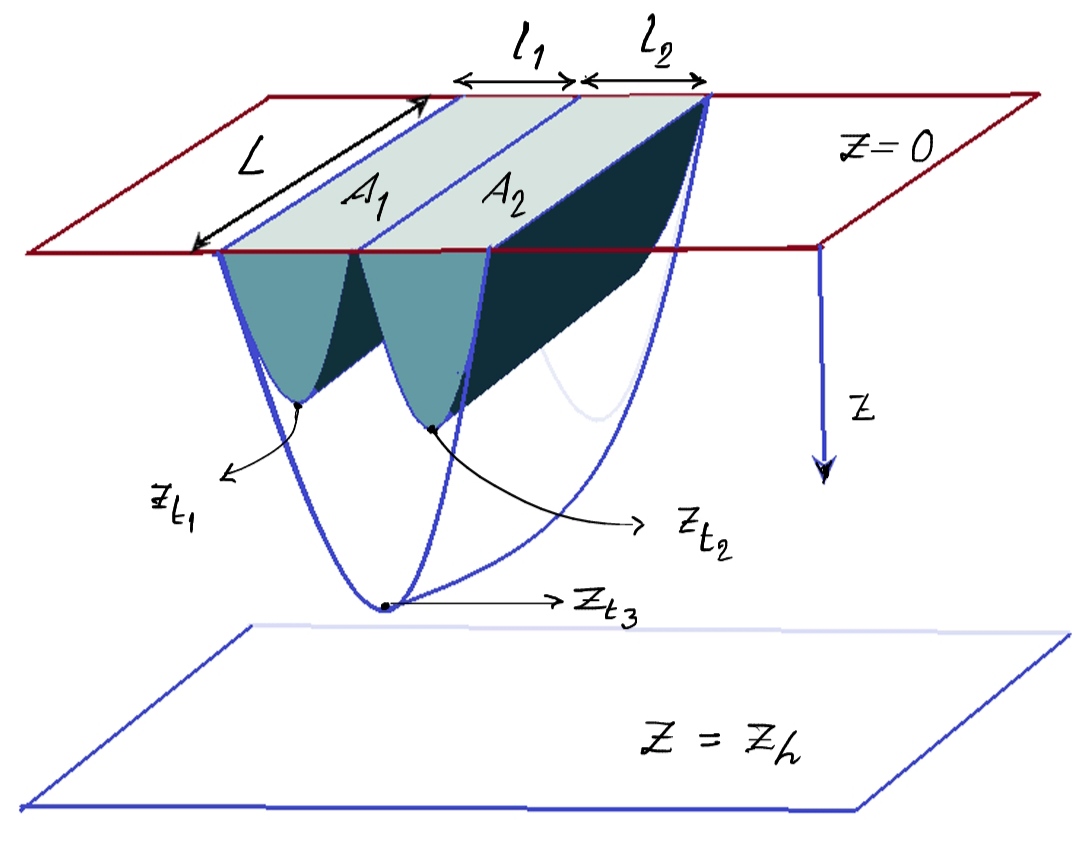}
\caption{Schematic diagram of the extremal surfaces, involving turning points, corresponding to two adjacent boundary subsystems $A$ and $B$ having widths $l_1$ and $l_2$ respectively. Here $z=0$ denotes the boundary whereas $z=z_h$ denotes the horizon. }
\label{adjlt}
\end{figure}

In this section, we utilize the holographic framework outlined in references \cite{Chaturvedi:2017znc, Chaturvedi:2016rft, Chaturvedi:2016rcn,  Jain:2017aqk} to analyze the holographic logarithmic negativity in the $1$RC black hole background. This calculation involves summing the areas of certain extremal surfaces in the bulk, associated with the relevant subsystems. As per the conjecture, the HLN for two adjacent subsystms can be expressed in the following manner,
\begin{equation}\label{eq19}
    \mathcal{E} = \frac{3}{16G_N^5}\left( \mathcal{A}_1 + \mathcal{A}_2 - \mathcal{A}_{12} \right),
\end{equation}
In this context, $\mathcal{A}_i$ represents the area of a co-dimension two extremal surface that is connected to the subsystem $A_i$ (refer to fig.\ref{adjlt}). It's worth noting that $\mathcal{A}_{12}$ specifically denotes the area of the extremal surface anchored to the combined subsystem $A_1\cup A_2$. As discussed in section \ref{sec:sec3}, we have shown the formula for the extremal surface's area related to a subsystem with a specific width. In the following subsections, we will apply these formulas to compute the HLN in both low and high-temperature scenarios.

\subsection{Holographic Logarithmic Negativity for two adjacent subsystems at low temperature}\label{subsec:sec4.1}

To begin with, we focus on the low-temperature regime 
and calculate the HLN for two neighboring subsystems of width $l_1$ and $l_2$. To check the  consistency of our findings, we take $\xi \to 0$ limit to reproduce the results for the AdS-Schwarzschild black hole as discussed in \cite{Jain:2017xsu}.

Before proceeding further, it is essential to acknowledge that in the low-temperature regime ($z_t/z_h \ll 1$), infinite series in both 
 equations \eqref{eq26} and \eqref{eq28} converge. Keeping this fact in mind we expand equation \eqref{eq28} up to the order of $z_t/z_h$ and obtain the following relation,
\begin{equation}\label{eq30}
l = z_t \Bigg{\{} a_1 - \frac{a_1\xi}{6} {\left( \frac{z_t}{z_h} \right)}^2 + \left[ \frac{a_2(1+\xi)}{2} + \frac{a_3 {\xi}^2}{24}\right] {\left(\frac{z_t}{z_h}\right)}^4 + \mathcal{O} {\left( \frac{z_t}{z_h}\right)}^6 \Bigg{\}}
\end{equation}
where the constants $a_1$, $a_2$ and $a_3$ are,
\begin{equation}\label{eq31}
\begin{split}
    & a_1 \equiv \sum_{j=0}^{\infty} \frac{\Gamma \left( j + \frac{1}{2} \right)}{\sqrt{\pi} \Gamma (j+1) \Gamma (2+3j)} = \frac{3\sqrt{\pi} \Gamma \left( \frac{5}{3}\right)}{\Gamma \left( \frac{1}{6}\right)} \\
    & a_2 \equiv \sum_{j=0}^{\infty} \frac{\Gamma \left( j + \frac{1}{2} \right)}{\sqrt{\pi} \Gamma (j+1) \Gamma (4+3j)} = \frac{\sqrt{\pi} \Gamma \left( \frac{7}{3}\right)}{4~\Gamma \left( \frac{11}{6}\right)} \\
    & a_3 \equiv \sum_{j=0}^{\infty} \frac{\Gamma \left( j + \frac{1}{2} \right) ( 4-j)}{\sqrt{\pi} \Gamma (j+1) \Gamma (2+3j) \Gamma (4+3j)} \\
    & = \frac{3}{\sqrt{\pi}} \left[ \Gamma \left( \frac{5}{6}\right)\Gamma \left( \frac{5}{3}\right) - \frac{3}{5} \Gamma \left( \frac{7}{6}\right) \Gamma \left( \frac{7}{3}\right) \right]- \frac{1}{70} ~{_3}F_2 \left( \frac{3}{2}, \frac{5}{3}, \frac{7}{3};\frac{8}{3},\frac{10}{3};1\right)
\end{split}
\end{equation}
Now it is straightforward to invert the relation between $l$ and $z_t$ as,
\begin{equation}\label{eq32}
z_{t}= \frac{l}{a_1} \Bigg{\{} 1 + \frac{\xi}{6a_1^2} {\left( \frac{l}{z_h}\right)}^2 + \frac{1}{24 a_1^4} \left[ \frac{{\xi}^2}{6}\left( 1 - \frac{a_3}{2a_2}\right) - \frac{a_2}{a_1}(1+\xi)\right] {\left( \frac{l}{z_h}\right)}^4 + \mathcal{O} {\left(  \frac{l}{z_h}\right)}^6\Bigg{\}}
\end{equation}
Similarly, we expand the infinite series for the area functional and get the following expression of area functional 
\begin{equation}\label{eq33}
\begin{split}
\mathcal{A}_{\text{finite}}^{\text{low}} &= \frac{L^2R^3}{z_t^2}\left[ \frac{1+\xi}{2}{\left( \frac{z_t}{z_h} \right)}^4 -1 \right] + \frac{L^2R^3}{z_t^2} \sum_{j=1}^{\infty} \frac{\Gamma \left( j + \frac{1}{2} \right)}{\sqrt{\pi} \Gamma (j+1) \Gamma (3j-1)} \\
& \times \left[ 1 + \frac{\xi}{3} {\left( \frac{z_t}{z_h} \right)}^2 + \left( \frac{(-4{\xi}^2 + 9\xi +9)j - 3(\xi +1)}{18j+6} \right){\left( \frac{z_t}{z_h} \right)}^4 \right]
\end{split}
\end{equation}
Now substituting the expression of the turning point \eqref{eq32} and performing the sum in \eqref{eq33} yields,
\begin{equation}\label{eq34}
\begin{split}
    \mathcal{A}_{\text{finite}}^{\text{low}} =\frac{R^3L^2}{l^2} \Bigg{\{} a_1^2 (w_1 -1) + \frac{\xi}{3}{\left( \frac{l}{z_h}\right)}^2 &+ \frac{1}{2a_1^2}\Bigg[ (1+\xi)\left(1-w_3+3w_2 + \frac{2(w_1-1)a_2}{a_1} \right)\\
    & +\frac{{\xi}^2}{6}\left(( w_1-1)(\frac{a_3}{a_1}-1)-8w_2\right)\Bigg] {\left( \frac{l}{z_h}  \right)}^4\Bigg{\}}
\end{split}
\end{equation}
where the numerical constants $w_1$, $w_2$, and $w_3$ are,
\begin{equation}\label{eq35}
\begin{split}
  & w_1\equiv \frac{1}{\sqrt{\pi}} \sum_{j=1}^{\infty} \frac{\Gamma \left( j+\frac{1}{2}\right)}{\Gamma (j+1)(3j-1)} = \frac{1}{2^{2/3}}~{_2}F_1\left( \frac{1}{3},\frac{2}{3};\frac{5}{3};-1 \right) \\
 & w_2\equiv \frac{1}{\sqrt{\pi}} \sum_{j=1}^{\infty} \frac{j \Gamma \left( j+\frac{1}{2}\right)}{\Gamma (j+1) (3j-1) (3j+1)} = \frac{1}{16}~{_3}F_2\left( \frac{2}{3},\frac{4}{3},\frac{3}{2};\frac{5}{3}, \frac{7}{3};1\right) \\
 & w_3\equiv \frac{1}{\sqrt{\pi}} \sum_{j=1}^{\infty} \frac{\Gamma \left( j+\frac{1}{2}\right)}{\Gamma (j+1) (3j-1) (3j+1)} = \frac{3}{16}~{_3}F_2\left( \frac{2}{3},\frac{4}{3},\frac{3}{2};\frac{5}{3}, \frac{7}{3};1\right) - \frac{1}{2^{1/3}}~{_2}F_1 \left( \frac{4}{3},\frac{5}{3};\frac{7}{3};-1\right)
\end{split}
\end{equation}
Note that in the limit where $\xi \to 0$, we get $z_h=1/\pi T$, and the subleading terms become $2^{\text{nd}}$ and $4^{\text{th}}$ order in $Tl$. To express this relation in a more simplified way, we define
\begin{equation}\label{eq36}
\begin{split}
    & c \equiv a_1^2(w_1 -1) \\
    & f(\xi) \equiv (1+\xi) b_1 + \frac{{\xi}^2}{6} b_2 \\
\end{split}
\end{equation}
\begin{equation}
    \text{Where,}~~
     b_1= \frac{\left(1-w_3+3w_2 + \frac{2(w_1-1)a_2}{a_1} \right)}{a_1^2}~~\text{and}~~b_2=\frac{\left(( w_1-1)(\frac{a_3}{a_1}-1)-8w_2\right)}{a_1^2}
\end{equation}
Using the above definitions we finally get the area functional at low temperature of the boundary subsystem in the form of a rectangular strip with width $l$ \cite{Ebrahim:2020qif} ,
\begin{equation}\label{eq37}
   \mathcal{A}_{\text{finite}}^{\text{low}} = R^3 {\left(\frac{L}{l}\right)}^2 \Bigg{\{} c + \frac{\xi}{3} {\left( \pi \hat{T}l\right)}^2 + \frac{1}{2} f(\xi) {\left( \pi \hat{T}l\right)}^4 \Bigg{\}}
\end{equation}
From  equations \eqref{eq36} and \eqref{eq37} it clear that as $\xi$ increases the area functional and therefore also the HEE  increase. In other words, the HEE shows an increment with respect to the charge of the black hole since parameter $\xi$ is proportional to the charge $Q$. A Similar characteristics of HEE for the charged AdS black hole is previously reported in \cite{Kundu:2016dyk}.
In the context of two adjoining subsystems, we designate these subsystems as $A_1$ and $A_2$, each defined by the width of their respective rectangular strips, $l_1$ and $l_2$ respectively. Using equation \eqref{eq19}, we derive the subsequent expression for the HLN within the low-temperature regime for the scenario involving two adjacent subsystems.
\begin{equation}\label{eq38}
\begin{split}
  \mathcal{E}_{low}= \frac{3R^3}{16G_N^5} \Bigg[ c\Bigg{\{} {\left(\frac{L}{l_1} \right)}^2 + {\left(\frac{L}{l_2} \right)}^2 - {\left(\frac{L}{l_1+l_2} \right)}^2\Bigg{\}} +\frac{\xi}{3}L^2{\pi}^2{\hat{T}}^2 - f(\xi)\left( \pi L^2 {\hat{T}}^4 \right) l_1 l_2 \Bigg]
  \end{split}
\end{equation}
The slop of HLN with respect to $\lambda$ can be written as,
 \begin{equation}\label{slop}
 \frac{d \mathcal{E}_{low}}{d \lambda}= \frac{d \mathcal{E}_{low}}{d \xi} \frac{d \xi}{d \lambda}
 \end{equation}
 From equations \eqref{eq36} and \eqref{eq38} we have,
\begin{equation}\label{slopadl}
\begin{split}
 \frac{d \mathcal{E}_{low}}{d \xi} = \frac{3R^3}{16G_N^5} \Bigg[ \frac{L^2{\pi}^2{\hat{T}}^2}{3} - \pi L^2\hat{T}^4 l_1 l_2 b_1+  \frac{(\pi L^2 {\hat{T}}^4  l_1 l_2 b_2)}{3} \xi \Bigg]
  \end{split}
\end{equation}
also from equation \eqref{xil},
\begin{equation}\label{dxil}
\frac{d \xi}{d \lambda}= \frac{4(1-\sqrt{1-\lambda^2})^2}{\lambda^3 \sqrt{1-\lambda^2}}
\end{equation}
Equation \eqref{slopadl} is finite in the critical limit $\xi \to 2$ while \eqref{dxil} is divergent at critical point $\lambda \to 1$ (equivalent to $\xi \to 2$) due to the factor $(1-\lambda^2)^{-\frac{1}{2}}$. Collectively the slope of HLN at low temperature limit diverges near the critical point with a power law behaviour which is similar to the power law  behaviour obtained for the slope of MI in \cite{Ebrahim:2020qif}.
Note that the explicit form of HLN (\ref{eq38}) is derived by considering the finite portion of the extremal areas of the adjacent subsystems.  
Nevertheless, if we take the complete area expression in to our consideration, the UV divergence term will also appear in the expression of negativity. We notice that all the $\xi$ dependent parts present in \eqref{eq38} collectively show an increasing behaviour with $\xi$. Also, as expected, in the critical limit $\xi \to 2$ the HLN is finite , since the individual HEEs are also finite at that parametric values. Now one can compare this HLN expression with the one in \cite{Jain:2017xsu} for the AdS$_{d+1}$ Schwarzschild black hole by setting $\xi \to 0$. The first three terms within the curly braces of \eqref{eq38} are inversely proportional to the squares of the lengths corresponding to the relevant boundary regions. The last term inside the square bracket depends on the product of the widths of two subregions.

\subsection{Holographic Logarithmic Negativity for two adjacent subsystems at high temperature}\label{subsec:sec4.2}
As mentioned in the previous section, there is always a concern regarding the divergence of infinite series. 
We observe that in the high-temperature limit $z_t \to z_h$, the infinite series in equation \eqref{eq25} does not exhibit convergence. Nevertheless, it is possible to regularize the series and extract the component proportional to $l$. In the limit $z_t \to z_h$, the expression for the area of the RT-surface takes the following form.
\begin{equation}\label{eq39}
\mathcal{A}_{\text{finite}}^{\text{high}} = R^3 {\left(\frac{L}{z_h}\right)}^2 \Bigg{\{} \sqrt{1+\xi}\left( \frac{l}{z_h} \right) + ( S_1 + S_2 +S_3) \Bigg{\}}
\end{equation}
where $S_1$, $S_2$ and $S_3$ dependent on $\xi $ and given by
\begin{equation}\label{eq40}
\begin{split}
& S_1 \equiv \frac{3\xi}{2} - \frac{1}{3} - \frac{11}{5\xi} - \frac{244}{105{\xi}^2}-\frac{32}{35{\xi}^3} - \frac{16}{35{\xi}^4} + \sqrt{\xi + 1} \left( -\frac{64\xi}{105}-\frac{124}{105}+\frac{26}{21\xi}+\frac{214}{105{\xi}^2}+ \frac{24}{35{\xi}^3} + \frac{16}{35{\xi}^4}\right) \\
& S_2 \equiv \sum_{k=2}^{\infty} \sum_{n=0}^{k} \sum_{m=0}^{\infty} \frac{\Gamma \left(k+\frac{1}{2}\right) \Gamma \left( m+\frac{1}{2}\right)\Gamma(k+n+2){\left(-1\right)}^{k+n}{\xi}^{k-n+m}{\left( 1+\xi\right)}^{n-m-\frac{1}{2}}}{\pi \Gamma(n+1)\Gamma(k-n+1)\Gamma(k+n+m+3)} \\
& \times \Bigg{\{} \frac{m+1}{k+n-1}\left[ 1 + \frac{m+1}{k+n} \left( 2 + \frac{m}{k+n+1}\right)\right] + \frac{(1+\xi)(m+1)}{k+n} \left( 2 + \frac{m}{k+n+1}\right) \Bigg{\}}
\end{split}
\end{equation}
\begin{equation} \label{eq41}
\begin{split}
& S_3 \equiv \sum_{k=2}^{\infty} \sum_{n=0}^{k} \sum_{m=0}^{\infty} \sum_{j=1}^{\infty} \frac{\Gamma \left(k+\frac{1}{2}\right) \Gamma \left( j+m+\frac{1}{2}\right)\Gamma(k+n+3j+2)}{\pi \Gamma(n+1)\Gamma(j+1)\Gamma(k-n+1)\Gamma(k+n+m+3j+3)}\\& \times {(-1)}^{k+n}{\xi}^{k-n+m}{(1+\xi)}^{n-m-\frac{1}{2}} \times \Bigg{\{} \frac{m+1}{k+n+3j-1} \left[ 1 + \frac{m+1}{k+n+3j}\left( 2 + \frac{m}{k+n+3j+1}\right) \right] \\& + \frac{(1+\xi)(m+1)}{k+n+3j}\left( 2 + \frac{m}{k+n+3j+1}\right)\Bigg{\}}
\end{split}
\end{equation}
Finally, in terms of temperature $\hat{T}$, we can write
\begin{equation}\label{eq42}
 \mathcal{A}_{\text{finite}}^{\text{high}} = R^3 {\left(\frac{L}{l}\right)}^2 \Bigg{\{} \sqrt{1+\xi} {\left( \pi \hat{T} l\right)}^3 + S_4 {\left( \pi \hat{T}l \right)}^2 \Bigg{\}}~~~\text{where}~~S_4 = S_1 + S_2 + S_3
\end{equation}
Therefore, using the formula for the HLN as given by equation \eqref{eq19} we find the HLN at a high-temperature regime for two adjacent subsystems $A_1$ and $A_2$ 
\begin{equation}\label{eq43}
  \mathcal{E}_{high}= \frac{3R^3}{16G_N^5} \Bigg{\{} S_4 L^2{\left(\pi \hat{T} \right)}^2 \Bigg{\}}
\end{equation}
 Slope of $\mathcal{E}_{high}$ with respect to $\lambda$ is given by ,
\begin{equation}\label{nhslop}
  \frac{d\mathcal{E}_{high}}{d\lambda}= \frac{3R^3}{16G_N^5} \Bigg{\{} {L\left(\pi \hat{T} \right)}^2 \frac{dS_4}{d \xi} \Bigg{\}} \frac{d\xi}{d\lambda}
\end{equation}
One can check that similar to the low temperature case the slope of $\mathcal{E}_{high}$ also shows a power law divergence due to the divergence of the quantity $\frac{d\xi}{d\lambda}$ at the critical point. 
Note that there is complete cancellation of the term proportional to $(Tl)^3$ in the HLN expression given by \eqref{eq43}. Therefore there is only single $T^2$ dependent term. As a consistency check we take  $\xi \to 0$ limit on the expression of HLN at high temperature for 1RCBH and reproduce the same for AdS$_{d+1}$ Schwarzschild black hole previously reported in \cite{Jain:2017xsu}. 
 In \eqref{eq42} the $\xi$ dependent is carried by the $S_4$. One can observe that in $\xi \to 2$ limit the HLN at high temperature is finite.

\section{Holographic Logarithmic Negativity for two disjoint subsystems}\label{sec:sec5}

\begin{figure}
\centering
 \includegraphics[width=.55\linewidth]{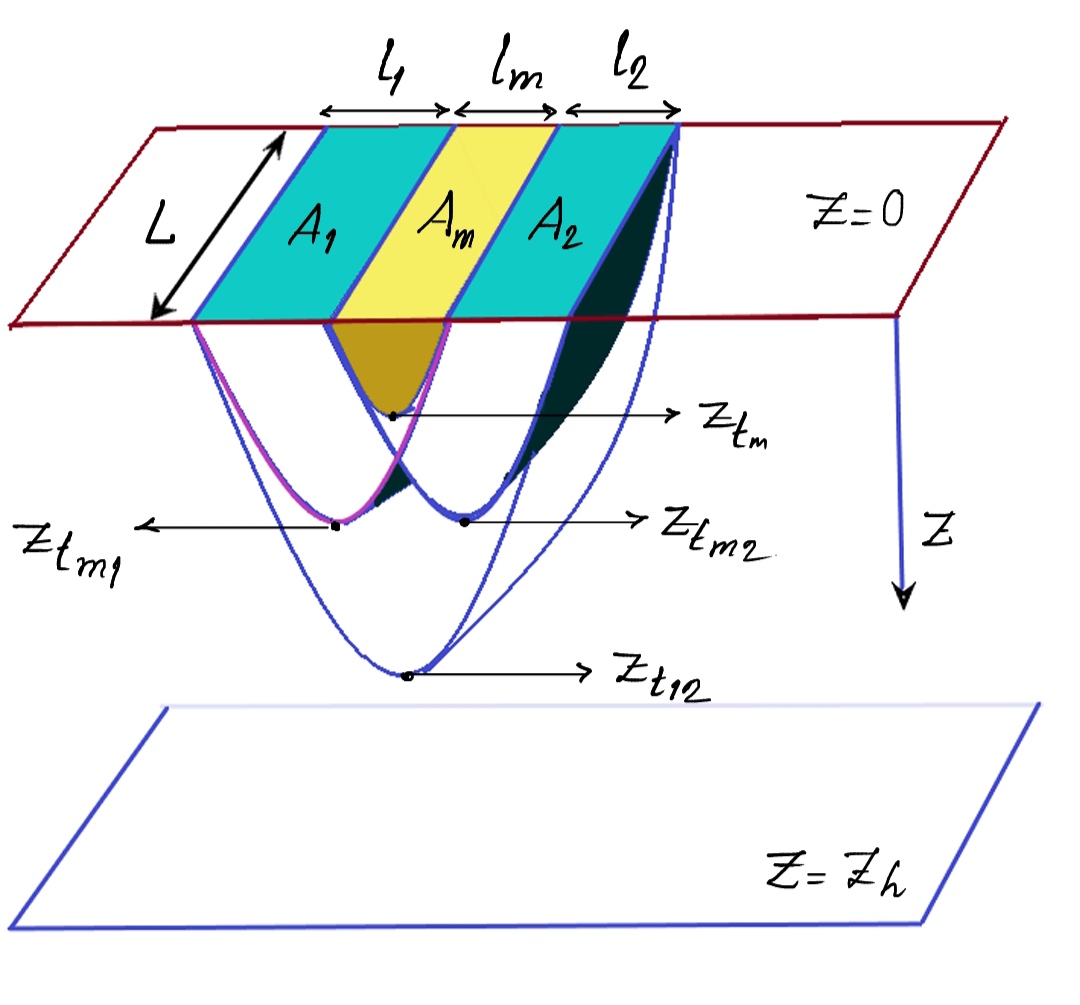}
\caption{Schematic diagram of the extremal surfaces at low effective temperature, involving the turning points, corresponding to the subregions $A_1$ and $A_2$ separated by an interval $A_m$. }
\label{dislt}
\end{figure}

In this section, we will determine the HLN for two disjoint subsystems in the strongly coupled field theory dual to the $1$RC black hole.   
Here we specifically focus on two non-overlapping intervals, denoted as $A_1$ and $A_2$, each with width $l_1$ and $l_2$ respectively, as illustrated in fig.\ref{dislt}. These intervals collectively constitute the subsystem $A$, with a separation $l_m$ corresponding to the width of a subsystem $A_m \subset B$ where $B = A^c$ represents the remainder of the system. To clarify, we define the three intervals as follows
\begin{equation}\label{eq44}
    A_1~:~~ x^1 \equiv x \in \left[-\frac{l_1}{2},\frac{l_1}{2}\right], \hspace{8mm} x^{(j)} \in \left[-\frac{L}{2},\frac{L}{2}\right]~~\text{where}~j=2, 3
\end{equation}
\begin{equation}\label{eq45}
     A_2~:~~  x^1 \equiv x \in \left[-\frac{l_2}{2},\frac{l_2}{2}\right], \hspace{8mm} x^{(j)} \in \left[-\frac{L}{2},\frac{L}{2}\right]~~\text{where}~j=2, 3
\end{equation}
\begin{equation}\label{eq46}
     A_m~:~~   x^1 \equiv x \in \left[-\frac{l_m}{2},\frac{l_m}{2}\right], \hspace{5mm} x^{(j)} \in \left[-\frac{L}{2},\frac{L}{2}\right]~~\text{where}~j=2, 3
\end{equation}
where $L$ is the length of all the transverse directions and it is taken to be very large $L \to \infty$.\\
Now following the conjecture given in \cite{KumarBasak:2020viv, Malvimat:2018txq} the entanglement negativity corresponding to the disjoint intervals becomes
\begin{equation}\label{eq47}
    \mathcal{E}=\frac{3}{16 G_N^5}\Big(\mathcal{A}_{A_1\cup A_m}+\mathcal{A}_{A_m\cup A_2}-\mathcal{A}_{A_1\cup A_m\cup A_2}-\mathcal{A}_{A_{m}}\Big)
\end{equation}
where $\mathcal{A}_{A_1\cup A_m}$, $\mathcal{A}_{A_m\cup A_2}$ and $\mathcal{A}_{A_{m}}$ are the areas of the extremal surfaces anchored with respect to the region $A_1 \cup A_m$, $A_2 \cup A_m$ and $A_m$ respectively. $\mathcal{A}_{A_1\cup A_m\cup A_2}$ is the area of the extremal surface anchored with respect to the region $A_1\cup A_m\cup A_2$. Note that the RT surfaces corresponding to the intervals $A_1$, $A_2$ and $A_m$ have the turning points $z_{t_1}$, $z_{t_2}$ and $z_{t_m}$ respectively.
By utilizing the area of the RT surfaces in their respective regions, we can calculate the HLN in the low and high-temperature limits.

\subsection{Holographic Logarithmic Negativity for two disjoint subsystems at low temperature}\label{sec5.1}

In our prior analysis, we have given the expression for the extremal surface area corresponding to a boundary region of width $l$ in the low-temperature limit \cite{Ebrahim:2020qif}.
\begin{equation}\label{eq48} 
   \mathcal{A}_{\text{finite}}^{\text{low}} = R^3 {\left(\frac{L}{l}\right)}^2 \Bigg{\{} c + \frac{\xi}{3} {\left( \pi \hat{T}l\right)}^2 + \frac{1}{2} f(\xi) {\left( \pi \hat{T}l\right)}^4 \Bigg{\}}
\end{equation}
Here we use the above relation to compute the expressions for the areas of the individual extremal surfaces associated with the intervals specified in \eqref{eq47}. The areas of the individual surfaces appearing in \eqref{eq47} are given by,

\begin{equation}\label{eq49}
\begin{split}
   & \mathcal{A}_{A_1\cup A_2 \cup A_m} = R^3 {\left(\frac{L}{l_1+l_2+l_m}\right)}^2 \Bigg{\{} c + \frac{\xi}{3} {\left( \pi \hat{T}(l_1+l_2+l_m)\right)}^2 + \frac{1}{2} f(\xi) {\left( \pi \hat{T}(l_1+l_2+l_m)\right)}^4 \Bigg{\}} \\
   &\mathcal{A}_{A_1 \cup A_m} = R^3 {\left(\frac{L}{l_1+l_m}\right)}^2 \Bigg{\{} c + \frac{\xi}{3} {\left( \pi \hat{T}(l_1+l_m)\right)}^2 + \frac{1}{2} f(\xi) {\left( \pi \hat{T}(l_1+l_m)\right)}^4 \Bigg{\}} \\
   &\mathcal{A}_{A_2 \cup A_m} = R^3 {\left(\frac{L}{l_2+l_m}\right)}^2 \Bigg{\{} c + \frac{\xi}{3} {\left( \pi \hat{T}(l_2+l_m)\right)}^2 + \frac{1}{2} f(\xi) {\left( \pi \hat{T}(l_2+l_m)\right)}^4 \Bigg{\}} \\
   &\mathcal{A}_{A_m} = R^3 {\left(\frac{L}{l_m}\right)}^2 \Bigg{\{} c + \frac{\xi}{3} {\left( \pi \hat{T}l_m\right)}^2 + \frac{1}{2} f(\xi) {\left( \pi \hat{T}l_m\right)}^4 \Bigg{\}}
\end{split}
\end{equation}
Using the above equation in \eqref{eq47} one would obtain the HLN at low temperatures for two disjoint subsystems
\begin{equation}\label{eq50}
\begin{split}
  \mathcal{E}_{low}= \frac{3R^3}{16G_N^5} \Bigg[ c\Bigg{\{} {\left(\frac{L}{l_1 + l_m} \right)}^2 + {\left(\frac{L}{l_2 + l_m} \right)}^2 - {\left(\frac{L}{l_1+l_2+l_m} \right)}^2 - {\left( \frac{L}{l_m} \right)}^2\Bigg{\}} \\ + \frac{1}{2}f(\xi)\left( \pi L^2 \hat{T}^4 \right) \Bigg{\{} {\left( l_1 + l_m \right)}^2 + {\left( l_2 + l_m \right)}^2 - {\left( l_1 + l_2 + l_m \right)}^2 - l_m^2 \Bigg{\}} \Bigg]
\end{split}
\end{equation}

Note that we are dealing exclusively with the finite portion of the area. This is why, in previous instances, no UV divergence term in HLN, $L^2/\epsilon^2$ are incorporated. However, in the scenario of disjoint intervals, a closer examination reveals that even if we consider the entire area expression, including the divergent part, the HLN remains independent of the cutoff. The first term on the right-hand side of \eqref{eq50} originates from the contribution of the AdS$_5$ vacuum and remains independent of the temperature. The remaining terms represent finite-temperature corrections to the HLN at low temperatures, closely resembling the conditions observed in the scenario of adjacent intervals. In the limit $\xi \to 0$, $\mathcal{E}_{low}$ correctly reduces to the previously documented result in \cite{KumarBasak:2020viv} for $AdS_{d+1}$ black hole. Since the $\xi$ dependent term in the above expression is contained in $f(\xi)$ which is finite in $\xi \to 2$ therefor $\mathcal{E}_{low}$ is also finite in this limit. We observed that the slope of \eqref{eq50} with respect to $\lambda$ diverges at critical point $\lambda\to 1$ and shows a power law divergence similar to the adjacent case.

 Additionally, one can anticipate that in the limit $l_m \to \epsilon$ the HLN for disjoint subsystems approaches to the result obtained for adjacent subsystems. Furthermore, a term reliant on the cutoff, $\frac{2}{d-2}{\left( \frac{L}{\epsilon}\right)}^{d-2}$ also emerges. Intriguingly, this term would have been present in the expression of HLN at low temperatures if the cutoff-dependent part within the region of the RT surfaces for multiple subregions had been considered. Hence, we can conclude that as $l_m$ approaches $\epsilon$, the HLN for separate subsystems converges to that of adjacent subsystems.

\subsection{Holographic Logarithmic Negativity for two disjoint subsystems at high temperature}\label{sec5.2}
In order to compute the HLN at high temperature we employ the area at high temperature \eqref{eq42} and write down the area of the extremal surfaces of all the intervals required in equation \eqref{eq47} as following,
\begin{equation}\label{eq52}
    \begin{split}
        &\mathcal{A}_{A_1\cup A_2\cup A_m}  = R^3 {\left(\frac{L}{l_1+l_2+l_m}\right)}^2 \Bigg{\{} \sqrt{1+\xi} {\left( \pi \hat{T} (l_1+l_2+l_m)\right)}^3 + S_4 {\left( \pi \hat{T}(l_1+l_2+l_m) \right)}^2 \Bigg{\}} \\
        &\mathcal{A}_{A_1\cup A_m} = R^3 {\left(\frac{L}{l_1+l_m}\right)}^2 \Bigg{\{} \sqrt{1+\xi} {\left( \pi \hat{T} (l_1+l_m)\right)}^3 + S_4 {\left( \pi \hat{T}(l_1+l_m) \right)}^2 \Bigg{\}} \\
        &\mathcal{A}_{A_2\cup A_m} = R^3 {\left(\frac{L}{l_2+l_m}\right)}^2 \Bigg{\{} \sqrt{1+\xi} {\left( \pi \hat{T} (l_2+l_m)\right)}^3 + S_4 {\left( \pi \hat{T}(l_2+l_m) \right)}^2 \Bigg{\}} \\
        &\mathcal{A}_{A_m} = R^3 {\left(\frac{L}{l_m}\right)}^2 \Bigg{\{} \sqrt{1+\xi} {\left( \pi \hat{T} l_m\right)}^3 + S_4 {\left( \pi \hat{T}l_m \right)}^2 \Bigg{\}}
    \end{split}
\end{equation}
Finally, following equation \eqref{eq47} the form of HLN at high temperature is given by ,
\begin{equation}\label{eq53}
\begin{split}
  \mathcal{E}_{high}=\frac{3R^3L^2}{16G_N^5}\Bigg{\{} \sqrt{1+\xi}{(\pi \hat{T})}^3(l_1+l_m) +S_4{(\pi \hat{T})}^2 + \sqrt{1+\xi}{(\pi \hat{T})}^3(l_2+l_m) +S_4{(\pi \hat{T})}^2 \\ - \sqrt{1+\xi}{(\pi \hat{T})}^3(l_1+l_2+l_m) -S_4{(\pi \hat{T})}^2 - \sqrt{1+\xi}{(\pi \hat{T})}^3 l_m -S_4{(\pi \hat{T})}^2 \Bigg{\}}=0
\end{split}
\end{equation}

The HLN vanishes at high temperature. This outcome aligns with expectations since logarithmic negativity exclusively quantifies quantum correlations, whereas in high temperature limit the dominant contribution comes from the thermal fluctuations on the boundary. A similar conclusion can be drawn for the observation reported in \cite{KumarBasak:2020viv} as demonstrated below,
\begin{figure}
\centering
 \includegraphics[width=.60\linewidth]{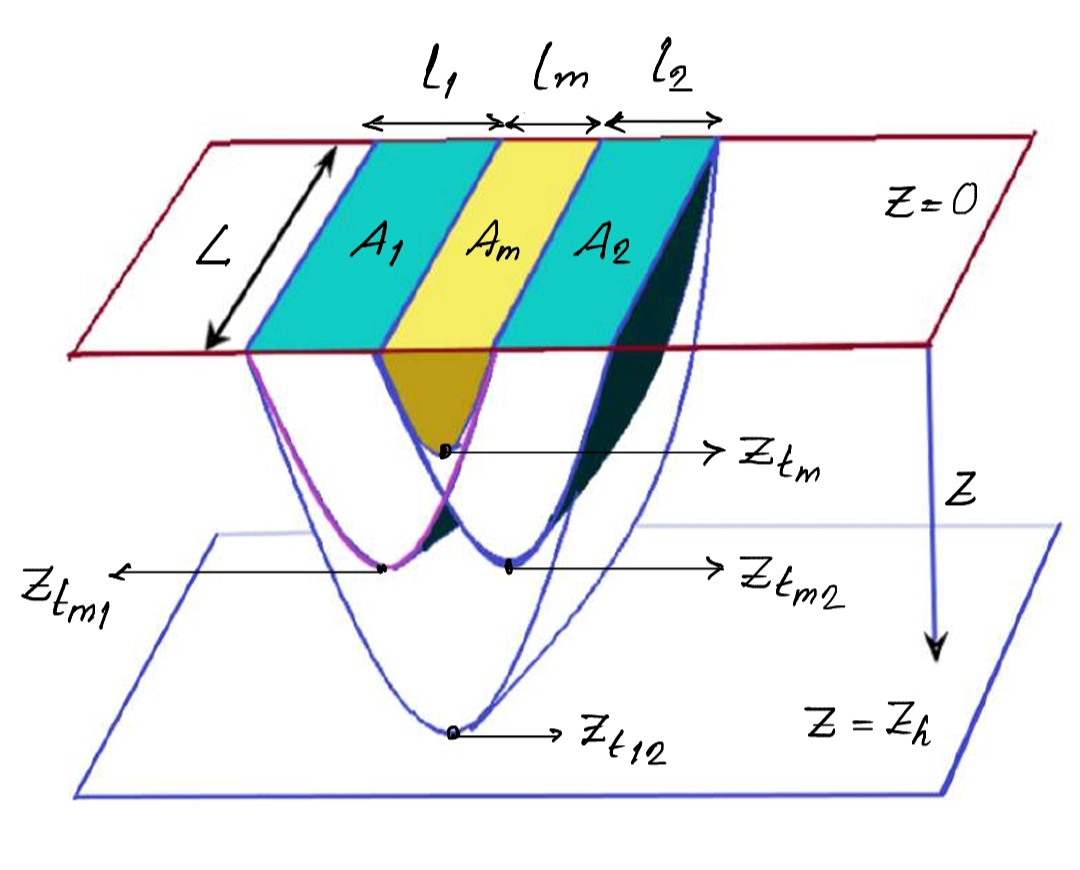}
\caption{Schematic diagram of the extremal surfaces at high effective temperature, involving the turning points, corresponding to the subregions $A_1$ and $A_2$ separated by an interval $A_m$.}
\label{disht}
\end{figure}

\begin{equation}\label{eq54}
\begin{split}
  \mathcal{E}=\frac{3}{16G_N^5}{\left(\frac{4\pi}{d}\right)}^{d-1}\frac{C_1}{4\pi}\sqrt{2d(d-1)}L^{d-2}T^{d-2} \Bigg{\{} - e^{-\sqrt{\frac{d-1}{2d}}4\pi T(l_1+l_m)} - e^{-\sqrt{\frac{d-1}{2d}}4\pi T(l_2+l_m)} \\ + e^{-\sqrt{\frac{d-1}{2d}}4\pi T(l_1+l_2+l_m)} + e^{-\sqrt{\frac{d-1}{2d}}4\pi T l_m} \Bigg{\}}
  \end{split}
\end{equation}
Above equation is the high temperature result for the $AdS_{d+1}$ black hole \cite{KumarBasak:2020viv}. In \ref{eq54} the exponential involves the product of temperature (T) and width ($l$) where $lT\gg1$.  Accordingly if one expands the exponential terms on the right-hand side of the above equation up to the linear order in $lT$ then the HLN vanishes.
\begin{equation}\label{eq55}
    \begin{split}
        \mathcal{E} = \frac{3}{16G_N^5}{\left(\frac{4\pi}{d}\right)}^{d-1}\frac{C_1}{4\pi}\sqrt{2d(d-1)}L^{d-2}T^{d-2} \Bigg{\{} -1+\sqrt{\frac{d-1}{2d}}4\pi T(l_1+l_m)-1\\ +\sqrt{\frac{d-1}{2d}}4\pi T(l_2+l_m) +1-\sqrt{\frac{d-1}{2d}}4\pi T(l_1+l_2+l_m) + 1- \sqrt{\frac{d-1}{2d}}4\pi T l_m \Bigg{\}=0}
    \end{split}
\end{equation}
This analysis confirm that our high-temperature HLN result for disjoint subsystems aligns with the findings in \cite{KumarBasak:2020viv}.

\section{Holographic Logarithmic Negativity for bipartite systems}\label{sec:sec6}

In this section, we calculate the HLN for a bipartite configuration for both low and high-temperature regimes.
 
To gain a clear understanding of this setup for the bipartite system, it is essential to begin by partitioning the boundary CFT into two subsystems, denoted as $A$ and its complement $A^c$. Furthermore, we  consider two additional subsystems, namely $B_1$ and $B_2$, located adjacently on the either sides of $A$ in such a way that we have $B = B_1 \cup B_2$.
 
The HLN for the bipartite system formed by the union of $A$ and $A^c$ is expressed as follows

\begin{equation}\label{eq56}
    \mathcal{E}=  \lim_{B\to A^c}\biggl[\frac{3}{16 G_N^{(d+1)}}\Big(2\mathcal{A}_{A}+\mathcal{A}_{B_1}+\mathcal{A}_{B_2}- \mathcal{A}_{A\cup B_1}-\mathcal{A}_{A\cup B_2}\Big)\biggr]
\end{equation}
where $G_N^{d+1}$ is the Newton's constant and $\mathcal{A}_i$ are the areas of the extremal surfaces. It is important to note that we can interpret the bipartite limit, denoted as $(B \to A^c)$, by extending the subsystems $B_1$ and $B_2$ to the extent such that $B$ effectively becomes equal to the complement of $A$. The subsystems of our interest, e.g.,  $A$, $B_1$, and $B_2$ are defined as:

\begin{equation}\label{eq57}
\begin{split}
   & A :~~ x^1 \equiv x \in \left[-\frac{l}{2},\frac{l}{2}\right], \hspace{9.5 mm} x^{(j)} \in \left[-\frac{L_2}{2},\frac{L_2}{2}\right]~~\text{where}~j=2, 3 \\
   & B_1 :~~ x^1 \equiv x \in \left[-L, -\frac{l}{2}\right], \hspace{5mm} x^{(j)} \in \left[-\frac{L_2}{2},\frac{L_2}{2}\right]~~\text{where}~j=2, 3 \\
   & B_2 :~~ x^1 \equiv x \in \left[\frac{l}{2}, L\right], \hspace{11.5 mm} x^{(j)} \in \left[-\frac{L_2}{2},\frac{L_2}{2}\right]~~\text{where}~j=2, 3
\end{split}
\end{equation}

 In this scenario, we can achive the bipartite limit by letting $B$ approach the complement of $A$, denoted as $B \to A^c$. This limit can be accommodated by treating the intervals $B_1$ and $B_2$ as infinitely large, i.e., $L\to \infty$. It is important to note that the choice of subsystem $A$ has been symmetrically made in such a way that $\mathcal{A}_{B_1} = \mathcal{A}_{B_2}$ and $\mathcal{A}_{A\cup B_1} = \mathcal{A}_{A\cup B_2}$. This identification is particularly elegant, as it simplifies the expression presented in equation \eqref{eq56} for the HLN to the following form
\begin{equation}\label{eq58}
    \mathcal{E}=\lim_{B\to A^c}\frac{3}{8 G_N^5}\Big(\mathcal{A}_A+\mathcal{A}_{B_1}- \mathcal{A}_{A\cup B_1}\Big)
\end{equation}
Here we designate the turning points associated with the RT surfaces anchored on regions $B_1$, $A$, and $A\cup B_1$ as $z_{t_1}$, $z_{t_2}$, and $z_{t_3}$ respectively. We utilize \eqref{eq58} to calculate the HLN in the low and high temperature scenario.

\subsection{Holographic Logarithmic Negativity for bipartite system at low temperature}\label{subsec:sec5.1}
In this section, we compute the HLN for the bipartite state in the low-temperature regime   
where we use the perturbative solution of the infinite series of $\frac{l}{2}$ as discussed in section \ref{sec:sec4}
\begin{figure}
\centering
 \includegraphics[width=.65\linewidth]{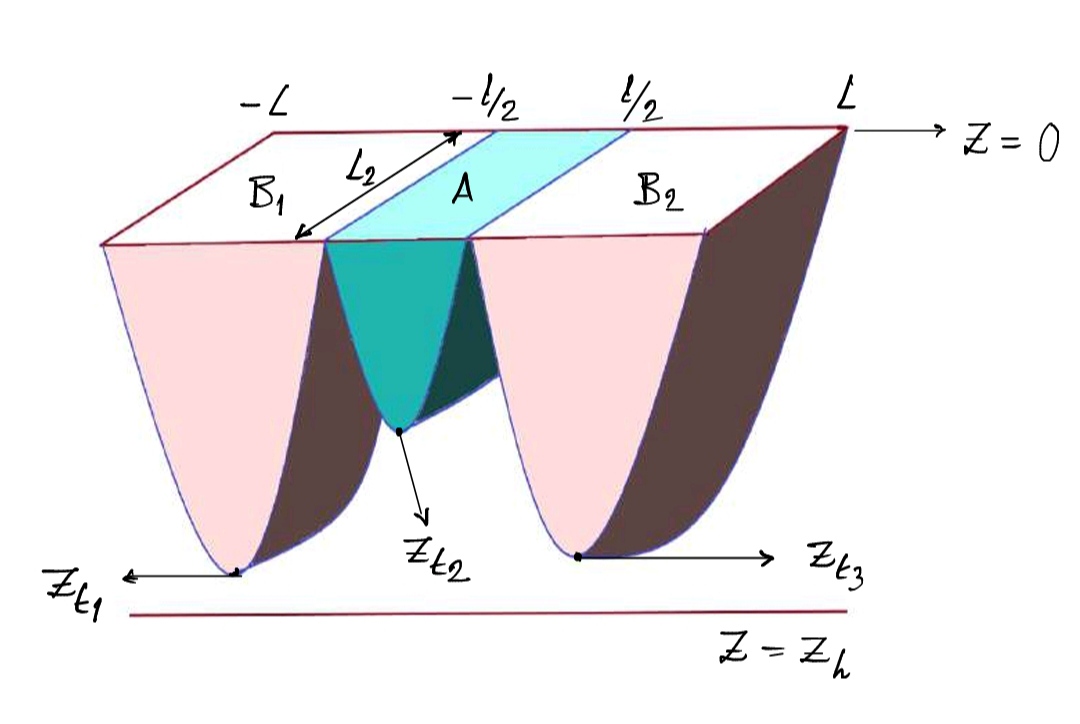}
\caption{Schematic diagram of the extremal surfaces corresponding to the bipartite subsystem at low effective temperature.}
\label{bplt}
\end{figure}

\begin{equation}\label{eq59}
z_{t_2}= \frac{l}{a_1} \Bigg{\{} 1 + \frac{\xi}{6a_1^2} {\left( \frac{l}{z_h}\right)}^2 + \frac{1}{24 a_1^4} \left[ \frac{{\xi}^2}{6}\left( 1 - \frac{a_3}{2a_2}\right) - \frac{a_2}{a_1}(1+\xi)\right] {\left( \frac{l}{z_h}\right)}^4 + \mathcal{O} {\left(  \frac{l}{z_h}\right)}^6\Bigg{\}}
\end{equation}
  The above relation along with the area functional give the low temperature expression for area of the extermal surface
 
\begin{equation}\label{eq60}
   \mathcal{A}_{A} = R^3 {\left(\frac{L}{l}\right)}^2 \Bigg{\{} c + \frac{\xi}{3} {\left( \pi \hat{T}l\right)}^2 + \frac{1}{2} f(\xi) {\left( \pi \hat{T}l\right)}^4 \Bigg{\}}
\end{equation}
The subsystems $B_1$ and $A \cup B_1$ in the boundary with lengths $(L - l/2)$ and $(L + l/2)$ along the $x^1$ direction becomes very large as $B \to A^c$ in the bipartite limit $ L \to \infty $. Since the boundary regions are very large the extremal surfaces described by the areas $\mathcal{A}_{B_1}$ and $\mathcal{A}_{A\cup B_1}$ extend deep into the bulk approaching the black hole horizon even at low temperatures i.e. $z_{t_1} \sim z_h$ and $z_{t_3} \sim z_h$. Hence, for computing the expressions for the areas $\mathcal{A}_{B_1}$ and $\mathcal{A}_{A\cup B_1}$ we employ the method developed in \cite{Fischler:2012ca} for the case when the RT surfaces approach the black hole horizon. We can write the turning points of the extremal surfaces $\mathcal{A}_{B_1}$ and $\mathcal{A}_{A\cup B_1}$  using the similar procedure for a $(d+1)$-dimensional AdS Schwarzchild black hole in \cite{Fischler:2012ca}.

\begin{equation}\label{eq61}
\begin{split}
    z_{t_1}=z_h(1+\epsilon_1)= z_h\Bigg[(1+k_2e^{-\sqrt{\frac{d(d-1)}{2}} z_h \left(L-\frac{l}{2}\right)}\Bigg] \\
    z_{t_3}=z_h(1+\epsilon_1)= z_h\Bigg[(1+k_2e^{-\sqrt{\frac{d(d-1)}{2}} z_h \left(L+\frac{l}{2}\right)}\Bigg]
\end{split}
\end{equation}
where $k_2$ has the following form
\begin{equation}\label{eq62}
    k_2 = \frac{1}{d}e^{\sqrt{\frac{d(d-1)}{2}}c_1}
\end{equation}
\begin{equation}\label{eq63}
    c_1 = \frac{2\sqrt{\pi}\Gamma\left( \frac{d}{2(d-1)}\right)}{\Gamma\left( \frac{1}{d-1} \right)} + \sum_{n=1}^{\infty} \Bigg{\{} \frac{2}{(1+nd)}\frac{\Gamma\left( n+ \frac{1}{2}\right)}{\Gamma(n+1)} \frac{\Gamma \left( \frac{d(n+1)}{2(d-1)}\right)}{\Gamma\left( \frac{nd+1}{2(d-1)} \right)} - \frac{\sqrt{2}}{\sqrt{d(d-1)}n} \Bigg{\}}
\end{equation}
We can now find out the area of the extremal surface for the subsystems $\mathcal{A}_{B_1}$ and $\mathcal{A}_{A\cup B_1}$ by substituting \eqref{eq61} in \eqref{eq25}. We then take the sum as an expansion of $\epsilon_1$ and $\epsilon_3$ respectively and consider the terms up to the linear order. 
\begin{equation}\label{eq64}
\begin{split}
    {\mathcal{A}}_{B_1} = \frac{L^2R^3}{z_h^2}\Big{\{} \alpha(\xi)+\gamma(\xi)+\mu(\xi) \Big{\}} + \frac{L^2R^3}{z_h}\left( L-\frac{l}{2}\right) \Big{\{}  \beta(\xi) +\delta(\xi) + \nu(\xi) \Big{\}} \\
     \mathcal{A}_{A\cup B_1} = \frac{L^2R^3}{z_h^2}\Big{\{} \alpha(\xi)+\gamma(\xi)+\mu(\xi) \Big{\}} + \frac{L^2R^3}{z_h}\left( L+\frac{l}{2}\right) \Big{\{}  \beta(\xi) +\delta(\xi) + \nu(\xi) \Big{\}}
\end{split}
\end{equation}
where all the $\xi$-dependent functions have been defined in Appendix- \ref{app:A}. Finally using equation \eqref{eq64} in \eqref{eq58} we achieve the form of HLN for the bipartite system at the low-temperature limit
\begin{equation}\label{eq65}
    \mathcal{E}_{low} =\frac{3}{8G_N^5} \Bigg[ R^3 {\left(\frac{L}{l}\right)}^2 \Bigg{\{} c + \frac{\xi}{3} {\left( \pi \hat{T}l\right)}^2 + \frac{1}{2} f(\xi) {\left( \pi \hat{T}l \right)}^4 \Bigg{\}} - R^3 L^2l\hat{T}g(\xi)\Bigg]
\end{equation}
where the function $g(\xi)$ can be written as $g(\xi) = \pi ( \beta(\xi) +\delta(\xi) + \nu(\xi))$. 
 From equation \eqref{eq65} the slope of $\mathcal{E}_{low}$ with respect to $\lambda$ is given by,
\begin{equation}\label{bils}
    \frac{d\mathcal{E}_{low}}{d\lambda}=\frac{d\mathcal{E}_{low}}{d\xi}\frac{d\xi}{d\lambda} =\frac{3}{8G_N^5} \Bigg[ R^3 {\left(\frac{L}{l}\right)}^2 \Bigg{\{} \frac{1}{3} {\left( \pi \hat{T}l\right)}^2 + \frac{1}{2} \frac{df(\xi)}{d\xi} {\left( \pi \hat{T}l \right)}^4 \Bigg{\}} - R^3 L^2l\hat{T}\frac{dg(\xi)}{d\xi}\Bigg] \frac{d\xi}{d\lambda}
\end{equation}
In the above equation the quantities $\frac{df(\xi)}{d\xi}$ and $\frac{dg(\xi)}{d\xi}$ are finite in critical limit while $\frac{d\xi}{d\lambda}$ diverges, as a result equation \eqref{bils} shows a power law divergence at $\lambda\to 1$.
 
In the expression of $\mathcal{E}_{low}$, the last term proportional to the $L^2l$ represents the three dimensional volume. One can interpret this term as the thermal entropy. In the $\xi \to 0$ limit the $\mathcal{E}_{low}$ reduces to the AdS Schwarzschild black hole result found in \cite{Chaturvedi:2016rft}. It is worth emphasizing that the initial term enclosed within the curly brackets, with the appropriate scaling factor, precisely reproduces the entanglement entropy of subsystem A at low temperatures.

In $\xi \to 0$ limit   $g(\xi)$ scales as $1/\xi$. Consequently, in terms of temperature, we can express this as $g(\xi) \propto T^2$. By combining these facts, it becomes apparent that the final term, which is proportional to the volume $V = L^2l$, exhibits an explicit temperature dependence of order $T^3$. Now we can reformulate equation \eqref{eq65} in the following manner

\begin{equation}\label{eq66}
    \mathcal{E}_{low}=\frac{3}{2} \Big{\{} S_A - \mathcal{C} S_A^{\text{Th}} \Big{\}},~~\text{where}~\mathcal{C}~\text{is a constant}
\end{equation}
 The above equation reveals that the HLN effectively quantifies distillable entanglement by eliminating the thermal contribution in low-temperature limit. This feature represents a universally observed characteristic of entanglement negativity in mixed states at finite temperatures.

\subsection{Holographic Logarithmic Negativity for Bipartite Systems at High Temperature}\label{subsec:sec5.2}

\begin{figure}
\centering
 \includegraphics[width=.65\linewidth]{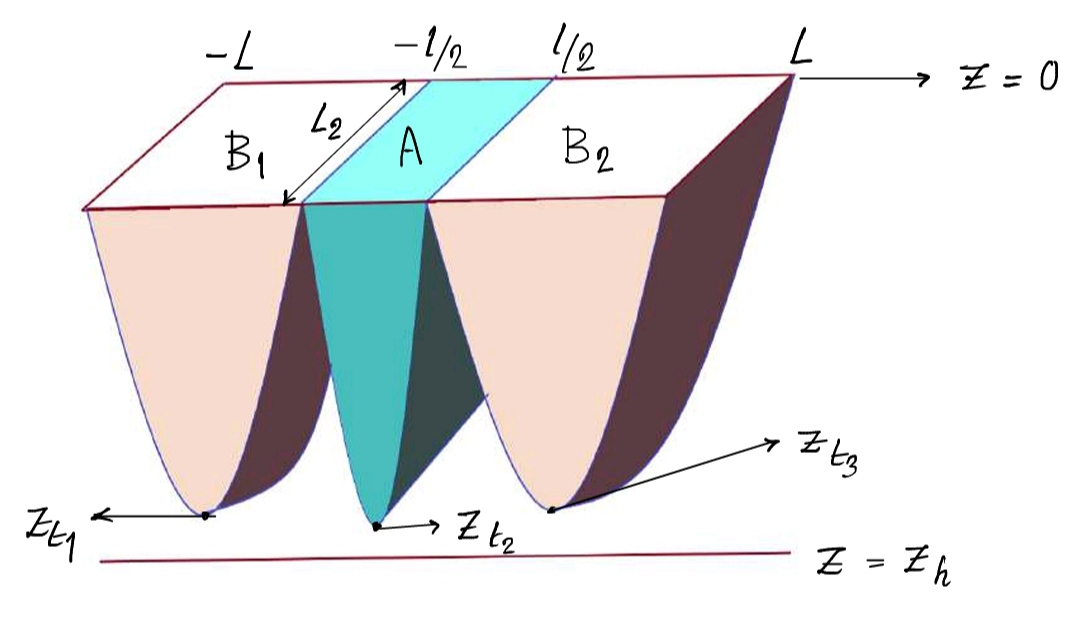}
\caption{Schematic diagram of the extremal surfaces corresponding to the bipartite subsystem at high effective temperature.}
\label{bpht}
\end{figure}

In the high-temperature regime, the turning point $z_{t_2}$ approaches to the black hole horizon, $z_{t_2}\sim z_h$( see fig. \ref{bpht}). Following the similar procedure as  discussed in the previous section \ref{subsec:sec5.1}, we can calculate the area of the extremal surfaces corresponding to the subsystem A. It is worth noting that, these surfaces extend into the vicinity of the black hole horizon, both at low and high temperatures. This behavior is a consequence of the limit $B \to A^c$, or equivalently, $L \to \infty$.
 
We write the following expression of the turning point corresponding to the extremal surface of the subsystem $A$

\begin{equation}\label{eq67}
    z_{t_2}=z_h(1+\epsilon_1)= z_h\Bigg[1+k_2e^{-\sqrt{\frac{d(d-1)}{2}} z_h l}\Bigg] 
\end{equation}
where $k_2$ is,
\begin{equation}\label{eq68}
    k_2 = \frac{1}{d}e^{\sqrt{\frac{d(d-1)}{2}}c_1}
\end{equation}
\begin{equation}\label{eq69}
    c_1 = \frac{2\sqrt{\pi}\Gamma\left( \frac{d}{2(d-1)}\right)}{\Gamma\left( \frac{1}{d-1} \right)} + \sum_{n=1}^{\infty} \Bigg{\{} \frac{2}{(1+nd)}\frac{\Gamma\left( n+ \frac{1}{2}\right)}{\Gamma(n+1)} \frac{\Gamma \left( \frac{d(n+1)}{2(d-1)}\right)}{\Gamma\left( \frac{nd+1}{2(d-1)} \right)} - \frac{\sqrt{2}}{\sqrt{d(d-1)}n} \Bigg{\}}
\end{equation}
The above equations for $d=4$ yields,
\begin{equation}\label{eq70}
    {\mathcal{A}}_{A} = \frac{L^2R^3}{z_h^2}\Big{\{} \alpha(\xi)+\gamma(\xi)+\mu(\xi) \Big{\}} + \frac{L^2R^3}{z_h}l \Big{\{}  \beta(\xi) +\delta(\xi) + \nu(\xi) \Big{\}} 
\end{equation}

Finally, by incorporating equations \eqref{eq70} and \eqref{eq64} into \eqref{eq58}, we arrive at the following expression for the HLN at high-temperature in the bipartite scenario,
\begin{equation}\label{eq71}
    \mathcal{E}_{high}= \frac{3}{8G_N^5}\Bigg[ \frac{L^2R^3}{z_h^2}\Big{\{} \alpha(\xi)+\gamma(\xi)+\mu(\xi) \Big{\}} + \frac{L^2R^3}{z_h}l \Big{\{}  \beta(\xi) +\delta(\xi) + \nu(\xi) \Big{\}} - L^2R^3 l\hat{T}g(\xi)\Bigg]
\end{equation}
One can obtain the slope of $\mathcal{E}_{high}$ with respect to $\lambda$ and can see that in this case also the power law divergent behaviour is true at the critical point $\lambda\to 1$.
Note that, as previously demonstrated for the low-temperature case, 
with the help of the limit defined as $\xi\to 0$, we can similarly reformulate the expression
for the HLN in a more concise way,

\begin{equation}\label{eq72}
    \mathcal{E}_{high}=\frac{3}{2} \Big{\{} S_A - \mathcal{C} S_A^{\text{Th}} \Big{\}},~~\text{where}~\mathcal{C}~\text{is a constant.}
\end{equation}

  Similar to the case of low temperature limit, in the high temperature regime the HLN picks up the contribution only from the distillable quantum entanglement.

\section{Entanglement Wedge Cross Section (EWCS)}\label{sec:sec7}

\begin{figure}
\centering
  \includegraphics[width=.60\linewidth]{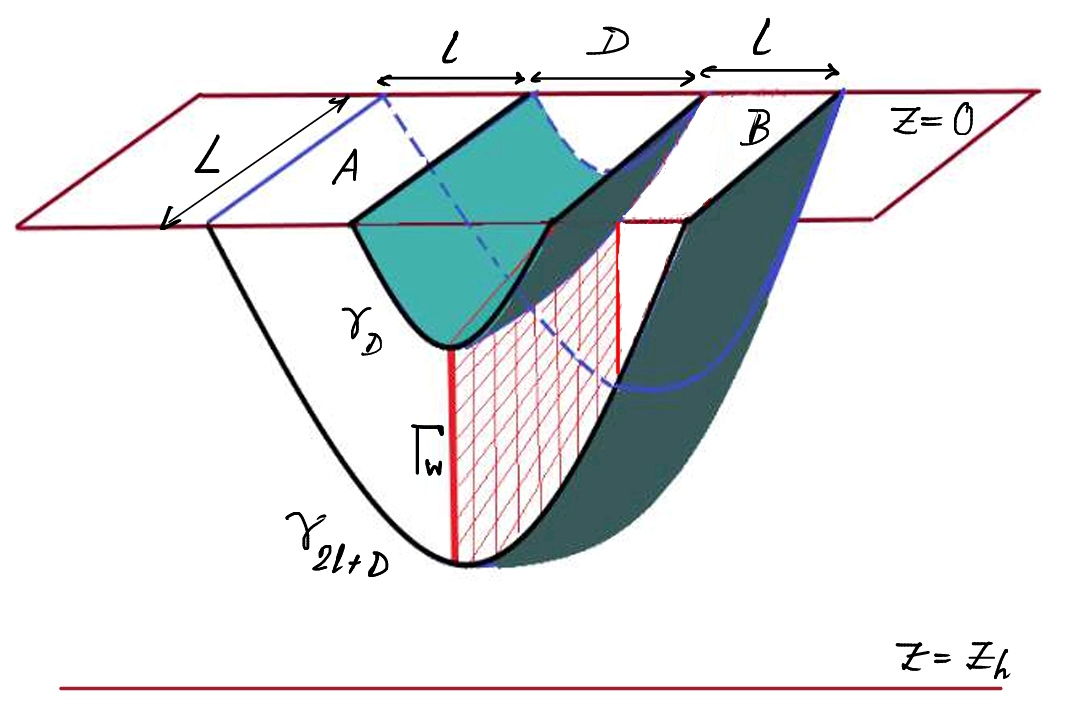}
  \caption{Schematic diagram of the extremal surfaces corresponding to two disjoint subsystems of equal length $l$ and separated by a distance $D$. The surface $\Gamma_W$, marked in red, is the entanglement wedge}
\label{ewcs}
\end{figure}

In this section, we compute the analytic form of the Entanglement Wedge Cross Section (EWCS) for low and high-temperature regimes. To describe the concept of the entanglement wedge, we take into account two subsystems $A$ and $B$ in the boundary. The Ryu-Takayanagi (RT) surface, represented as $\gamma_{AB}$, characterizes the region encompassing $A\cup B$. The entanglement wedge is defined as the volume enclosed by the boundary $A\cup B\cup \gamma_{AB}$ and EWCS is the extremal area surface $\Gamma_W$, which bifurcates that volume, as illustrated in fig. \ref{ewcs}.

 In \cite{Amrahi:2020jqg}, an exploration of EWCS for 1RC black hole background is accomplished by using the numerical analysis. In our present analysis, we derive an analytic expression for the EWCS for the same black hole background. In the present context, we define the boundary subsystems $A$ and $B$, each with a length of $l$ and separated by a distance $D$ in the following way,
\begin{equation}\label{ewcs1}
\begin{split}
 & A :~~ x^1 \equiv x \in \left[-l-\frac{D}{2},-\frac{D}{2}\right], \hspace{5mm} x^{(j)} \in \left[-\frac{L_2}{2},\frac{L_2}{2}\right]~~\text{where}~j=2, 3 \\
   & B :~~ x^1 \equiv x \in \left[l+\frac{D}{2},\frac{D}{2}\right], \hspace{12 mm} x^{(j)} \in \left[-\frac{L_2}{2},\frac{L_2}{2}\right]~~\text{where}~j=2, 3 
\end{split}
\end{equation}

 In this setup $\Sigma_{\text{min}}$ is the surface with minimum area positioned at $x=0$ that separates the subsystem $A$ and $B$. The induced metric on this surface is given by:

\begin{equation}\label{ewcs2}
    ds_{\Sigma_{min}}^2 = e^{2A(z)}d{\vec{x}}_{2}^2 + \frac{e^{2B(z)}}{h(z)}\frac{R^4}{z^4}dz^2
\end{equation}
  From the above metric one can compute the EWCS as
\begin{equation}\label{ewcs3}
    E_W = \frac{L^{2}}{4G_N^{5}} \int_{z_t(D)}^{z_t(2l+D)}dz \sqrt{g_{mn}}
\end{equation}
where $g_{\text{mn}}$ is the induced metric. $z_t(2l+D)$ and $z_t(D)$ are the turning points of the extremal surfaces we have considered. Therefore equation \eqref{ewcs2} and \eqref{ewcs3} gives,
\begin{equation}\label{ewcs4}
     E_W = \frac{L^{2}R^2}{4G_N^5} \int_{z_t(D)}^{z_t(2l+D)}{dz \frac{e^{2A(z)+B(z)}}{z^2 \sqrt{h(z)}}}
\end{equation}
By employing \eqref{eq4} and the definition of the dimensionless parameter $\xi$, we can express the above integral as follows, \footnote{Note that in deriving equation \eqref{ewcs5}, we have applied the multinomial expansion, similar to our earlier approach in section \ref{sec:sec4}.}

\begin{equation}\label{ewcs5}
\begin{split}
    E_W = \frac{L^2R^3}{4G_N^5}\int_{z_t(D)}^{z_t(2l+D)} dz \sum_{k=0}^{\infty} \sum_{j=0}^{k} \sum_{i=0}^{\infty} \frac{{(-1)}^{k+j}}{2}\frac{\Gamma(k+\frac{1}{2}) {\xi}^{i+j+k}{(1+\xi)}^j}{\Gamma(i+1)\Gamma(\frac{3}{2}-i)\Gamma(j+1)\Gamma(k-j+1)} \frac{z^{2i+2j+2k-3}}{{z_h}^{2i+2j+2k}}\\
    = \frac{L^2R^3}{4G_N^5} \sum_{k=0}^{\infty} \sum_{j=0}^{k} \sum_{i=0}^{\infty} \frac{{(-1)}^{k+j}}{2}\frac{\Gamma(k+\frac{1}{2}) {\xi}^{i+j+k}{(1+\xi)}^j}{\Gamma(i+1)\Gamma(\frac{3}{2}-i)\Gamma(j+1)\Gamma(k-j+1)}\frac{1}{(2i+2j+2k-2)}\\ 
    \times \Bigg{\{} \frac{z_t(2l+D)^{2i+2j+2k-2}}{z_h^{2i+2j+2k}} - \frac{z_t(D)^{2i+2j+2k-2}}{z_h^{2i+2j+2k}}\Bigg{\}}
\end{split}
\end{equation} 
As indicated by the expression above, it becomes evident that when $D$ significantly exceeds $l$, the EWCS completely disappears.
In the following subsections, we  examine the behavior of the EWCS at low and high temperature.

\subsection{Entanglement Wedge Cross Section at low temperature}\label{subsec:sec8.1}

Since it is generally challenging to invert the relation between $z_t$ and width $l$ and achieve an expression for EWCS in terms of the boundary parameters for an arbitrary temperature, we resort to low and high-temperature limits. In the low-temperature limit, where $z_t(D)\ll z_H$ and $z_t(2l+D)\ll z_H$, we use equation \eqref{eq30} and derive the following expressions for the turning points.

\begin{equation}\label{ewcs6}
z_{t}(D)= \frac{D}{a_1} \Bigg{\{} 1 + \frac{\xi}{6a_1^2} {\left( \frac{D}{z_h}\right)}^2 + \frac{1}{24 a_1^4} \left[ \frac{{\xi}^2}{6}\left( 1 - \frac{a_3}{2a_2}\right) - \frac{a_2}{a_1}(1+\xi)\right] {\left( \frac{D}{z_h}\right)}^4 + \mathcal{O} {\left(  \frac{D}{z_h}\right)}^6\Bigg{\}}
\end{equation}
\begin{equation}\label{ewcs7}
\begin{split}
z_{t}(2l+D)= \frac{2l+D}{a_1} \Bigg{\{} 1 + \frac{\xi}{6a_1^2} {\left( \frac{2l+D}{z_h}\right)}^2 + \frac{1}{24 a_1^4} \left[ \frac{{\xi}^2}{6}\left( 1 - \frac{a_3}{2a_2}\right) - \frac{a_2}{a_1}(1+\xi)\right] {\left( \frac{2l+D}{z_h}\right)}^4 \\ + \mathcal{O} {\left(  \frac{2L+D}{z_h}\right)}^6\Bigg{\}} 
\end{split}
\end{equation}

Finally by substituting equations \eqref{ewcs6} and \eqref{ewcs7} into equation \eqref{ewcs5}, we derive an analytical  expression for EWCS in the low-temperature regime. Note that we can use the binomial expansion to both the turning points in order to simplify the EWCS at low temperature limit. This expansion is feasible because the coefficients of $\mathcal{O}(1/z_h^2)$, $\mathcal{O}(1/z_h^4)$, etc., within the parentheses in equations \eqref{ewcs6} and \eqref{ewcs7}, are smaller than unity in the low-temperature limit.
Further simplifications in the resulting equation (see Appendix \ref{app:B}) can be achieved by truncating the series at the lowest order, which corresponds to setting $i=j=k=0$. It is also reasonable to assert that since at low temperatures, both $D$ and $l$ are small it allows us to neglect higher order terms associated with these length scales. We can express the simplified version of the EWCS at low temperatures as follows.

In terms of temperature the above expression becomes
\begin{equation}\label{ewcs9}
    E_W^{\text{low}} = \frac{L^2R^3}{4G_N^5} \Bigg[ \frac{a_1^2}{2} \Bigg{\{} \frac{1}{D^2} - \frac{1}{{(2l+D)}^2}\Bigg{\}} + \frac{2}{a_1^2}\Bigg{\{} \frac{{\xi}^2}{6}\left( 1- \frac{a_3}{a_2}\right)-\frac{a_2}{a_1}(1+\xi)\Bigg{\}}l(l+D){(\pi \hat{T})}^4 + \mathcal{O}\left({\hat{T}}^6\right) \Bigg]
\end{equation}
 In the expression of $E_W^{\text{low}}$ the first term enclosed in the curly braces is independent of temperature. This term implies that as the separation between subsystems decreases the EWCS increases and in the limit where D approaches zero, it becomes unbounded.

Using the connection between the EWCS and mutual information as discussed in \cite{Umemoto:2019jlz}, we can observe that, at low temperatures, the EWCS exhibits a similar behavior to that of mutual information. In the critical limit, denoted as $\xi \to 2$, EWCS remains finite. Similar characteristic is observed for mutual information in \cite{Ebrahim:2020qif}. The power law divergent behaviour of the MI near the critical point obtained in \cite{Ebrahim:2020qif} can also be seen by the slope of equation \eqref{ewcs9} as,
\begin{equation}\label{ewsll}
    \frac{dE_W^{\text{low}}}{d\lambda} = \frac{dE_W^{\text{low}}}{d\xi} \frac{d\xi}{d\lambda}
\end{equation}
From equation \eqref{ewcs9} it is clear that $\frac{dE_W^{\text{low}}}{d\xi}$ is finite in the limit $\xi \to 2$ or $\lambda\to 1$ and from equation \eqref{dxil} the quantity $\frac{d\xi}{d\lambda}$ is divergent at $\lambda \to 1$. As a result the slope of $E_W^{\text{low}}$ shows a power law divergence near the critical point $\lambda\to 1$. 

\subsection{Entanglement Wedge Cross Section at High Temperature}\label{subsec:sec8.2}

We now examine the EWCS at high temperatures. The analytic form of EWCS can be achieved by following two viable approaches based on the choice of the boundary parameter $l$ and $D$. In the first approach $D$ is chosen to be very large but finite and both the turning points corresponding to the extremal surfaces $\gamma_D$ and $\gamma_{2l+D}$  move deeper into the bulk. As a result, one can, in principle, employ the near-horizon expansion for the turning points $z_t(D)$ and $z_t(2l+D)$. However, in this set up, as $D$ tends to infinity, the EWCS at high temperatures diminishes.
In the second approach one can consider $l$ approaching to infinity while $D$ is held fixed at a small value. In this scenario, a non-zero, significantly large value for the EWCS is expected, and this can be obtained by focusing on the near-horizon expansion for the extremal surface $\gamma_{2l+D}$. For our purpose we follow the first approach.

Utilizing the techniques discussed in the previous sections, we find with the following expression for the turning point,

\begin{equation}\label{ewcs10}
    z_t(D)=z_h(1+\epsilon) = z_h\left(1+k_2e^{-\sqrt{\frac{d(d-1)}{2}}z_h D}\right)
\end{equation}
where,
\begin{equation}\label{ewcs11}
    k_2 = \frac{1}{d}e^{\sqrt{\frac{d(d-1)}{2}}c_1}
\end{equation}
\begin{equation}\label{ewcs12}
    c_1 = \frac{2\sqrt{\pi}\Gamma\left( \frac{d}{2(d-1)}\right)}{\Gamma\left( \frac{1}{d-1} \right)} + \sum_{n=1}^{\infty} \Bigg{\{} \frac{2}{(1+nd)}\frac{\Gamma\left( n+ \frac{1}{2}\right)}{\Gamma(n+1)} \frac{\Gamma \left( \frac{d(n+1)}{2(d-1)}\right)}{\Gamma\left( \frac{nd+1}{2(d-1)} \right)} - \frac{\sqrt{2}}{\sqrt{d(d-1)}n} \Bigg{\}}
\end{equation}
Note that we are working with $d=4$. We insert the the expression of $z_t$ given in \eqref{ewcs10} in the last line of \eqref{ewcs5} and obtain 
\begin{equation}\label{ewcs13}
\begin{split}
    E_W^{\text{high}}  = \frac{L^2R^3}{4G_N^5} \sum_{k=0}^{\infty} \sum_{j=0}^{k} \sum_{i=0}^{\infty} \frac{{(-1)}^{k+j}}{2}\frac{\Gamma(k+\frac{1}{2}) {\xi}^{i+j+k}{(1+\xi)}^j}{\Gamma(i+1)\Gamma(\frac{3}{2}-i)\Gamma(j+1)\Gamma(k-j+1)}\frac{1}{(2i+2j+2k-2)} \\ 
    \times \Bigg{\{} \frac{{\left( 1 +  k_2 e^{-\sqrt{6}z_h(2l+D)}\right)}^{2i+2j+2k-2}}{z_h^{2}} - \frac{{\left( 1 +  k_2 e^{-\sqrt{6}z_h D }\right)}^{2i+2j+2k-2}}{z_h^{2}}\Bigg{\}}
\end{split}
\end{equation}
Taking the binomial expansion in the above equation up to order $\epsilon$ and by suppressing the higher order terms EWCS at high temperature takes the form as,
\begin{equation}\label{ewcs14}
    E_W^{\text{high}}  = \frac{L^2R^3}{4G_N^5} \sum_{k=0}^{\infty} \sum_{j=0}^{k} \sum_{i=0}^{\infty} \frac{{(-1)}^{k+j}}{2}\frac{\Gamma(k+\frac{1}{2}) {\xi}^{i+j+k}{(1+\xi)}^j}{\Gamma(i+1)\Gamma(\frac{3}{2}-i)\Gamma(j+1)\Gamma(k-j+1)}\frac{k_2}{z_h^2} \Bigg( e^{-\sqrt{6}z_h(2l+D)} - e^{-\sqrt{6}z_hD} \Bigg)
\end{equation}
In terms of temperature, we can rewrite the above expression in the following way,
\begin{equation}\label{ewcs15}
    E_W^{\text{high}}  = \frac{L^2R^3}{4G_N^5} \sum_{k=0}^{\infty} \sum_{j=0}^{k} \sum_{i=0}^{\infty} \frac{{(-1)}^{k+j}}{2}\frac{\Gamma(k+\frac{1}{2}) {\xi}^{i+j+k}{(1+\xi)}^j}{\Gamma(i+1)\Gamma(\frac{3}{2}-i)\Gamma(j+1)\Gamma(k-j+1)}{(\pi \hat{T})}^2  \Bigg( e^{-\frac{\sqrt{6}(2l+D)}{\pi \hat{T} }} - e^{-\frac{\sqrt{6} D}{\pi \hat{T}}} \Bigg)
\end{equation}
Similar to the previous section, we will now refer to the calculation of mutual information at high temperatures as presented in \cite{Ebrahim:2020qif}. In equation \eqref{ewcs15}, we can apply the high-temperature limit by considering $D$ to be large but finite. In the Case $D\gg l$ the EWCS at high temperature becomes zero. This is consistent with the findings of \cite{Ebrahim:2020qif}, where mutual information vanishes for the large separation between the subsystems. It's worth noting,  that working with the non-trivial limit $l\gg D$ in the boundary parameter could be an interesting exercise, demonstrating that this limit corresponds to a substantial EWCS value. However, we reserve this exploration for future research. Similar to the low temperature case a power law divergent  is also followed by the slope of the $E_W^{\text{high}}$ at the critical point.

\section{Holographic Thermo Mutual Information (HTMI)}\label{sec:sec8}

The holographic dual of a thermofield double (TFD) state is essential for studying information scrambling in the context of AdS/CFT correspondence. TFD state is a pure entangled state that belongs to a bipartite Hilbert space made out of the individual Hilbert spaces of two identical copies of field theory.

The holographic dual of the TFD state is two sided eternal black hole. 
\begin{figure}
\centering
  \includegraphics[width=.40\linewidth]{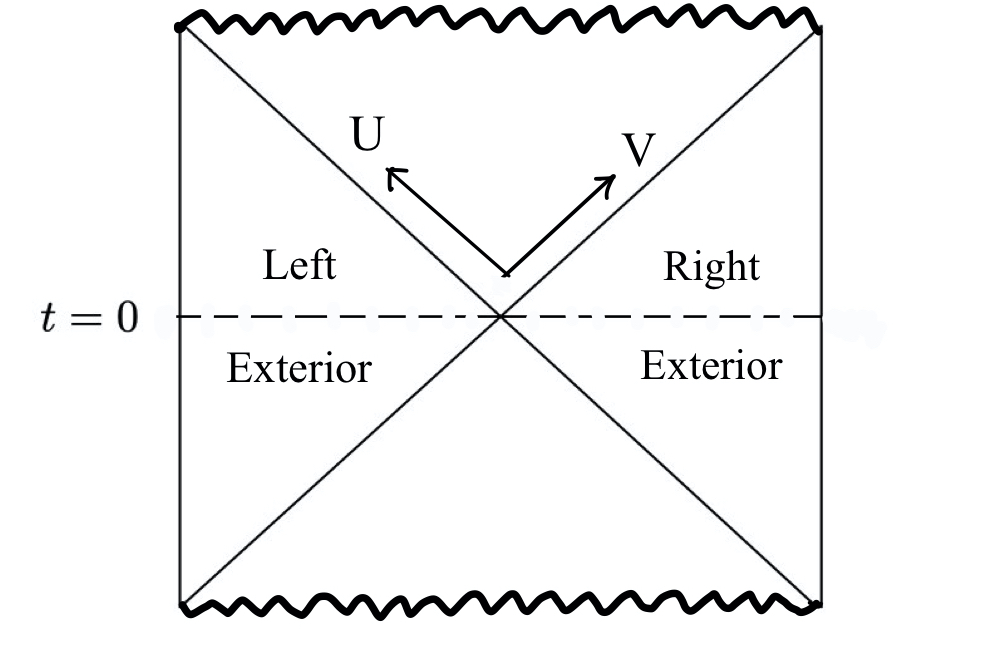}
  \caption{ Penrose diagram of the eternal black hole. At t = 0, the spatial extremal surface connecting two asymptotic boundaries of an eternal black hole is denoted by the dashed line passing through the bifurcation point.
  }
\label{penrose}
\end{figure}
 The region outside the horizon of eternal black hole is made out of two causally disconnected parts named as left (L) and right (R) wedges respectively. It is possible to have a non-local correlation between two boundary theories that are separately residing in the asymptotic boundaries of the L and R regions respectively.
Thermo mutual information (TMI) is an appropriate measure to study such correlations first introduced in \cite{Morrison:2012iz} along with its holographic dual known as holographic thermo mutual information (HTMI).   
  The HTMI later studied in \cite{Leichenauer:2014nxa,Chakrabortty:2022kvq,Jahnke:2017iwi} and the references therein. One can weaken these correlations by applying a small perturbation that grows exponentially with time. Hollographically this perturbation can be interpreted as an in going shockwave in the dual bulk spacetime.

\subsection{Holographic Thermo Mutual Information (HTMI)}\label{subsec:sec7.1}

To determine the HTMI, we adopt the methodology presented in \cite{Jahnke:2017iwi,Chakrabortty:2022kvq, Leichenauer:2014nxa}. Consider two subsystems A and B each of width $l$. The HTMI is given by $I(A:B)=\mathcal{S}(A)+\mathcal{S}(A)-\mathcal{S}(A\cup B)$. Subsystem $A$ is positioned along the left asymptotic boundary, while subsystem $B$ is situated along the right asymptotic boundary of the eternal black hole at time $t=0$. As per the RT proposal, $\mathcal{S}(A)$ and $\mathcal{S}(B)$ are directly linked to the minimal surface areas of $\gamma_A$ and $\gamma_B$ corresponding to the entangling regions $A$ and $B$ respectively.

We define the embedding for $\gamma_i (i = A, B)$ as ($t=0, z, -l/2 \leq x (z) \leq l/2, -L/2 \leq x^j \leq L/2 $ $j=2,3$).
 To find $\mathcal{S}(A\cup B)$ we have two different choices of the extremal surface, of which one is $\gamma_A \cup \gamma_B$ and the other is $ \gamma_1 \cup \gamma_2 = \gamma_{\text{wormhole}}$. For positive HTMI we choose the second one i.e, $ \gamma_1 \cup \gamma_2$. Surfaces  $ \gamma_1$ and $ \gamma_2$ are connecting two asymptotic boundaries through the bifurcation point of the eternal black hole denoted by the doted line in the Fig.\ref{penrose}. The appropriate embedding for $\gamma_1$ and $\gamma_2$ are ($t=0, z,  x=-l/2, -L/2 \leq x^j \leq L/2$) and ($t=0, z,  x=l/2, -L/2 \leq x^j \leq L/2$) respectively.
HTMI becomes zero as $\mathcal{A}(\gamma_A \cup \gamma_B) \leq \mathcal{A}(\gamma_{\text{wormhole}})$, and positive for the opposite situation. To find area of $\gamma_{\text{wormhole}}$ we use the prescription given by Hubeny-Rangamani-Takayanagi (HRT)\cite {Hubeny:2007xt}. The components of the induced metrics on the corresponding RT and HRT surface are respectively given as, 
\begin{equation}\label{eq74}
    G_{\text{in}}^A= G_{\text{in}}^B={L^2e^{3A(z)}}\left(\frac{R^4 e^{2(B(z)-A(z))}}{z^4h(z)}+x'^2\right)^{\frac{1}{2}}
,
\qquad
    G_{\text{in}}^{\text{wormhole}}
=
    \frac{L^2R^2e^{B(z)+2A(z)}}{z^2\sqrt{h(z)}}
,
\end{equation}
where $x'=0$ for $\gamma_1$ and $\gamma_2$. Now, HTMI is
\begin{equation}\label{eq75}
\begin{split}
    I(A,B)
&= 
   \frac{1}{4 G_N^5} 
    \left[  
    2
    \int_{0} ^{z_t}
    dz
    \left(
    \sqrt{ G_{in}^{A} } 
    +
    \sqrt{G_{in}^{B}}
    \right)
    -  
    4
    \int_{0} ^{z_h}
    dz
    \sqrt{G_{in}^{A\cup B}}
    \right]
\\   
&=
    \frac{
    L^{2}R^{2}}{G_N^5}
    \left[
    \int_{0}^{z_t}\frac{e^{B(z)+2A(z)}dz}{z^{2}\sqrt{h(z)}}\sqrt{(\frac{e^{6A(z_t)}}{e^{6A(z)}-e^{6 A(z_t)}}+1)}-\int_{0}^{z_h}\frac{e^{B(z)+2A(z)}dz}{z^{2}\sqrt{h(z)}}
    \right]
\end{split}
\end{equation}

 The parameter $z_t$ denotes the turning point of the RT surfaces associated with regions $A$ and $B$.
\begin{figure}
\centering
 \includegraphics[width=.50\linewidth]{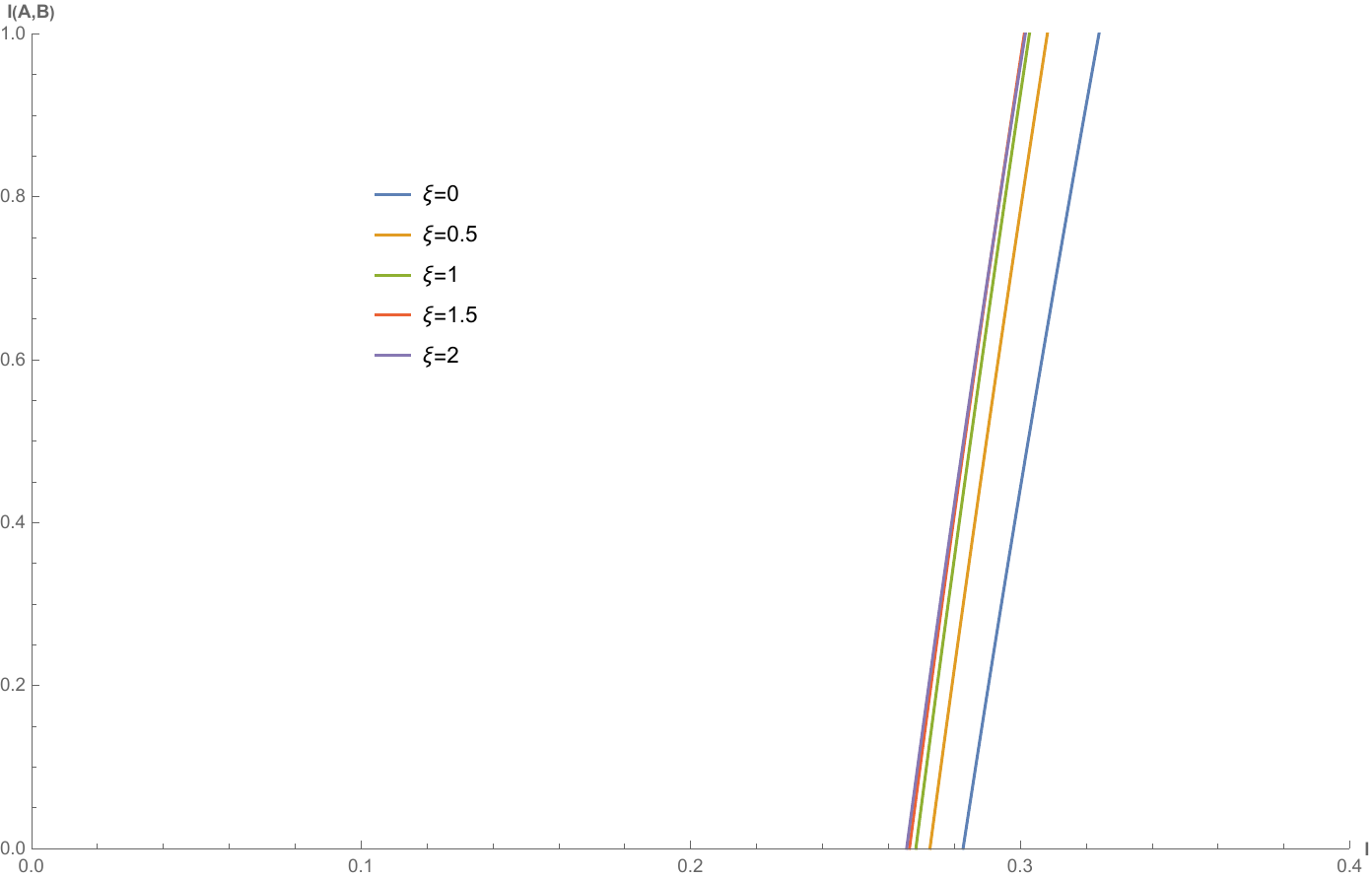}
\caption{HTMI with respect to width $l$ for $T=1, R=1$ with different values of $\xi$}
\label{TMIl}
\end{figure}
The relationship between the HTMI and the width ($l$) of entangling region can be determined through the use of the $l$ and $z_t$ relation.
\begin{equation}\label{eq76}
 \frac{l}{2}=\int_{0}^{z_t}\frac{dz}{\sqrt{\left(\frac{Q^2 z^2}{R^4}+1\right) \left(\frac{z_t^6 \left(\frac{Q^2
   z^2}{R^4}+1\right)}{z^6 \left(\frac{Q^2 z_t^2}{R^4}+1\right)}-1\right) \left(1-\frac{z^4
   \left(\frac{Q^2 z_h^2}{R^4}+1\right)}{z_h^4 \left(\frac{Q^2 z^2}{R^4}+1\right)}\right)}}
\end{equation} 
The Fig.\ref{TMIl} illustrates that, as the width $l$ increases, the HTMI also takes higher values. However, when we reach a specific value of $l$ (say critical width $l_c$), the HTMI is zero for any $l\leq l_c$. This critical value of width decreases as we raise the value of $\xi$. This is because, when $l \leq l_c$, the HRT surface corresponding to $A\cup B$ accumulates greater area than the combined areas of the individual RT surfaces associated with $A$ and $B$. Consequently, the selection for the $A\cup B$ surface will be equivalent to the sum of the areas of the RT surfaces corresponding to $A$ and $B$. Also note that as the value of $\xi$ increases
 the critical width $l_c$ takes the smaller value. As $\xi$ approaches the critical point of the theory $\xi \to 2$, the two consecutive $l_c$ comes closer. The increasing behaviour of HTMI with respect to the width of the subregion is previously reported in different scenarios \cite{ Jahnke:2017iwi, Chakrabortty:2022kvq}.

\subsection{Holographic Thermo Mutual Information with shockwave}\label{subsec:sec7.2}

\begin{figure}
\centering
 \includegraphics[width=.50\linewidth]{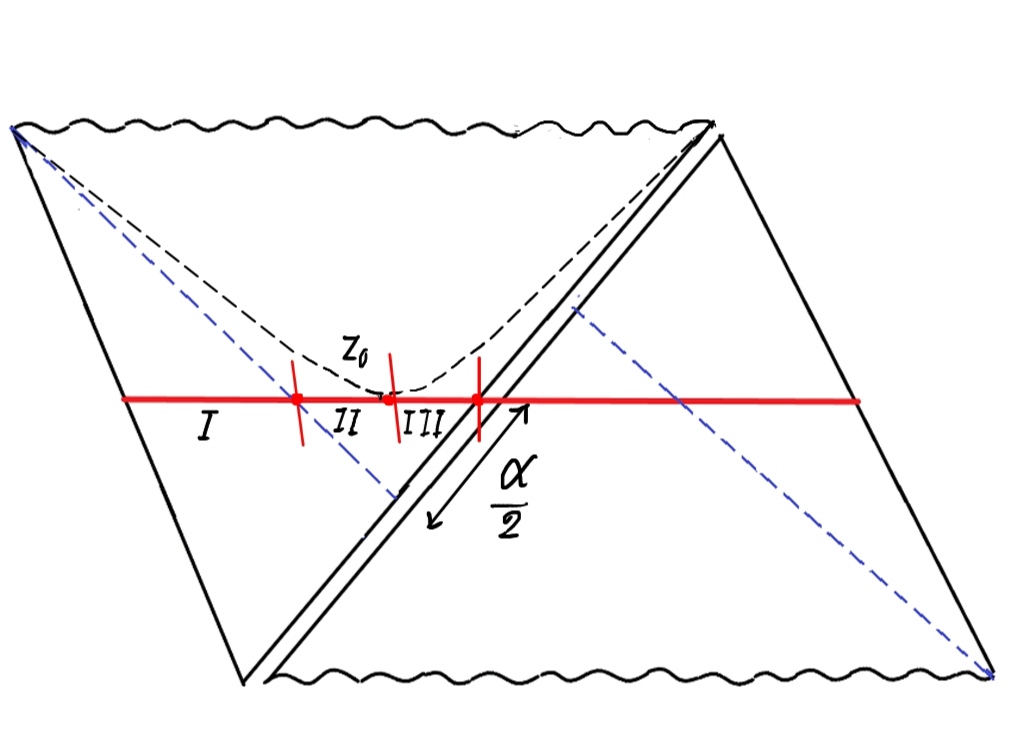}
\caption{Penrose diagram after the shock wave. Red line shows the wormhole surface with turning point $z_0$ connecting $A$ and $B$.}
\label{PDshock}
\end{figure}

 In the holographic setup we consider that a tiny energy pulse is injected from the left boundary in the asymptotic past towards the bulk region. As time passes this tiny pulse of energy grows exponentially due to the blue shift and becomes a shockwave that backreacts the geometry. The impact of the shock wave on the geometry can be accounted by adjusting the Kruskal coordinate from $V$ to $\hat{V}=V+\Theta(U)\alpha$, while leaving all other coordinates unchanged and denoting them with a hat, as previously demonstrated in \cite{Leichenauer:2014nxa,Chakrabortty:2022kvq,Jahnke:2017iwi}. The function $\Theta(U)$ ensures that the shockwave's influence is confined to the left region of the Penrose diagram shown in Fig. \ref{penrose}, which is modified as depicted in Fig.\ref{PDshock}.

To remove the divergences of the HTMI due to the presence of shockwave we follow the definition 

\begin{equation}\label{eq77}
 I(A:B;\alpha)= I(A,B;\alpha=0)- \mathcal{S}^{\text{reg}}_{A\cup B}(\alpha),
\end{equation}
where $I(A: B; \alpha = 0)$ is the TMI for $\alpha=0$ previously calculated in equation \eqref{eq75}, and $\mathcal{S}^{\text{reg}}_{A\cup B}(\alpha)= \mathcal{S}_{A\cup B}(\alpha)- \mathcal{S}_{A\cup B}(\alpha=0)$.
is the regularized entropy.

 To find $\mathcal{S}^{\text{reg}}_{A\cup B}(\alpha)$ we choose a set of time-dependent embeddings defined as $\{t, z(t),  x=-l/2, -L/2 \leq x^j \leq L/2\}$ and $\{t, z(t),  x=l/2, -L/2 \leq x^j \leq L/2\} $. The area functional corresponding to either of these time-dependent embeddings is given as,
\begin{equation}\label{eq78}
\mathcal{A}=L^2 \int{dt\biggl[-e^{6A(z)}h(z) +\dot{z}^2\frac{R^4 e^{2B(z)+4A(z)}}{z^4h(z)})\biggr]^{\frac{1}{2}}},~~ \mathcal{L}= L^2\biggl[{-e^{6A(z)}h(z) +\dot{z}^2\frac{R^4 e^{2B(z)+4A(z)}}{z^4h(z)}\biggr]^{\frac{1}{2}}} 
\end{equation}
Note that the Lagrangian density $\mathcal{L}$ does not contains the time explicitly which leads to the conserved quantity, $\mathcal{P}=-L^2 e^{3A(z_0)}\sqrt{-h(z_0)}$ based on boundary conditions $\Dot{z}=0|_{z=z_0}$. Now, we can write,
\begin{equation}\label{eq79}
     \dot{z}^2=\frac{z^2 h(z)}{R^4 e^{4A(z)+2B(z)}}\biggl[\frac{L^4 h^2(z) e^{12A(z)}}{\mathcal{P}}+e^{6A(z)} h(z)\biggr]^{\frac{1}{2}}
\end{equation}
Substituting equation \eqref{eq79} in \eqref{eq78} the area functional,
  \begin{equation}\label{eq80}
     \mathcal{A}=L^2 R^2 \int_{0}^{z_0}dz~{\frac{ e^{5A(z)+B(z)}}{z^2\sqrt{\mathcal{P}^2+L^4h(z)e^{6A(z)}}}}
  \end{equation}
Also by integrating equation \eqref{eq79} we get,
\begin{equation}\label{eq81}
    t(z)=\pm \int dz~ {\frac{ R^2 e^{B(z)-A(z)}}{z^2 h(z)\sqrt{(1+\frac{L^4 h(z) e^{6A(z)}}{\mathcal{P}^2})}}}
\end{equation}
Using the conserved momenta $\mathcal{P}$ in equation \eqref{eq80} and \eqref{eq81} yields,
\begin{equation}\label{eq82}
 \mathcal{A}=L^2 R^2 \int_{0}^{z_0}dz~{\frac{ e^{5A(z)+B(z)}}{z^2\sqrt{h(z)e^{6A(z)}-h(z_0)e^{6A(z_0)}}}},~~ t(z)=\pm \int dz~ {\frac{ R^2 e^{B(z)-A(z)}}{z^2 h(z)\sqrt{1-\frac{ h(z)}{h(z_0)}e^{6A(z)-6A(z_0)}}}}.
\end{equation}

\begin{figure}
\centering
 \includegraphics[width=.50\linewidth]{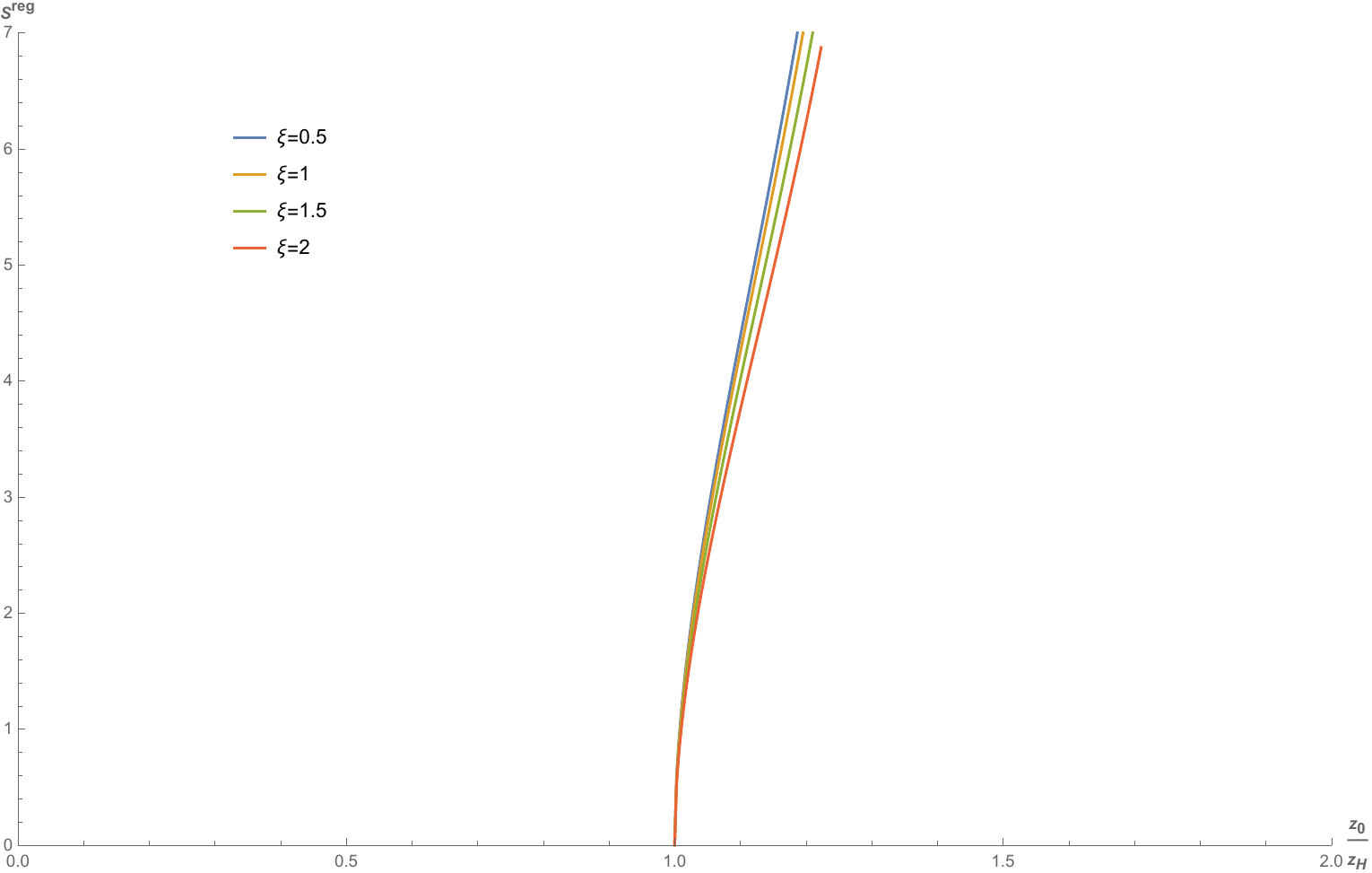}
 \caption{$S^{reg}$ v/s $\frac{z_0}{z_H}$ for different $\xi$ with $T=1, R=1$.}
\label{srgvsa}
\end{figure}
 For the calculation of $\mathcal{A}(\gamma_{\text{w}})$, we split the integration $\mathcal{A}$  into three distinct regions. The region (I) starts at the left boundary and spans inside the bulk upto the left horizon. The second region (II) starts from the horizon and ended at $z=z_0$. Lastly, the third region (III) begins at $z=z_0$, proceeds towards the right (as depicted in Fig.\ref{PDshock}), and extends until it encounters the right horizon. The sign ``$\pm$" signifies the rate of change of $z$ with respect to $t$ along the extremal surface.

Taking into account the three regions labeled as I, II, and III, the expression for $\mathcal{A}(\gamma_{\text{w}})$ takes the following form: 
\begin{equation}\label{eq83}
\begin{split}
\mathcal{A}(\gamma_{\text{w}}) = 
4 L^2 R^2 & \biggl[\int^{z_H}_{0}{dz\biggl(\frac{e^{5A(z)+B(z)}}{z^2\sqrt{h(z)e^{6A(z)}-h(z_0)e^{6A(z_0)}}}}-\frac{ e^{2A(z)+B(z)}}{z^2 \sqrt{h(z)}}\biggr)\\&+2 \int^{z_0}_{z_H}{dz\frac{e^{5A(z)+B(z)}}{z^2\sqrt{h(z)e^{6A(z)}-h(z_0)e^{6A(z_0)}}}}\biggr]
\end{split}
\end{equation}

The regularized HEE is expressed as $S^{\text{reg}}{A\cup B}(z_0) = \frac{\mathcal{A}(\gamma_{w})}{4 G_N^5}$. This regularized entropy is depicted in Fig.\ref{srgvsa} as a function of the dimensionless parameter $\frac{z_0}{z_h}$. It is evident from the plot that as the ratio $z_0/z_h$ increases from unity, the value of $S^{\text{reg}}$ initiates from zero at $z_0=z_h$ and progressively rises. The rate of change of regularized entropy with respect to $z_h/z_0$ is more when the parameter $\xi$ is kept at a fixed smaller value.

By employing equations \eqref{eq77} and \eqref{eq83}, we can express the HTMI as a function of $z_0$. To examine how the HTMI changes concerning the shock wave parameter $\alpha$, we must establish the connection between $z_0$ and $\alpha$.
In Fig.\ref{PDshock}, Region (I) is defined as the area between the boundary point $(\hat{U},\hat{V})=(1,-1)$ and the point on the horizon $(\hat{U},\hat{V})=(\hat{U}_1,0)$. Region (II) spans from $(\hat{U},\hat{V})=(\hat{U}_1,0)$ to a point denoted as $(\hat{U},\hat{V})=(\hat{U}_2,\hat{V}_2)$, while Region (III) extends from $(\hat{U},\hat{V})=(\hat{U}_2,\hat{V}_2)$ to $(\hat{U},\hat{V})=(0,\alpha/2)$.

Using the definition of Kruskal coordinates, it is possible to express the variation of  $\hat{U}=\pm e^{\frac{2\pi}{\beta}(z_*-t)}$ and $\hat{V}=\pm e^{\frac{2\pi}{\beta}(z_*+t)}$ as, 
\begin{equation}\label{eq84}
\begin{split}
&\Delta \log \hat{U}^2=\log \hat{U}_1^2-\log{\hat{U}_0^2} =\frac{4\pi}{\beta}(\Delta z_*-\Delta t)\\& \Delta \log\hat{V}^2=\log \hat{V}_2^2-\log{\hat{V}_1^2}=\frac{4\pi}{\beta}(\Delta z_*+\Delta t)
\end{split}
\end{equation}

\begin{equation}\label{eq85}
\begin{split}
&\log{\hat{U}}=\frac{2\pi}{\beta}\int dz~{\frac{R^2 e^{B(z)-A(z)}}{z^2 h(z)}\Bigg(\frac{1}{\sqrt{1-\frac{h(z)}{h(z_0)}e^{6A(z)-6A(z_0)}}}-1\Bigg)} \\
& \log{\hat{V} }=\frac{2\pi}{\beta}\int dz~{\frac{R^2 e^{B(z)-A(z)}}{z^2 h(z)}\Bigg(\frac{1}{\sqrt{1-\frac{h(z)}{h(z_0)}e^{6A(z)-6A(z_0)}}}+1\Bigg)}
\end{split}
\end{equation}
where $z_*$ is defined as follows, 
\begin{equation}\label{eq86}
    z_*=-\int dz~ {\frac{R^2e^{B(z)-A(z)}}{z^2h(z)}}
\end{equation}

Note that in region (I), when $\dot{z}<0$, it leads to an overall negative sign in the expression for $t$. Conversely, in region (II), the negative numerical value of $h(z)$ corresponds to $\dot{z}>0$, and hence we introduce a negative sign. Now, let's consider the variation of $\hat{U}$ from the boundary to the horizon.
\begin{equation}\label{eq87}
    \hat{U}^2_1=\exp{\Bigg[\frac{4\pi}{\beta}\int_{0}^{z_H}dz~{\frac{R^2 e^{B(z)-A(z)}}{z^2 h(z)}\Bigg(\frac{1}{\sqrt{1-\frac{h(z)}{h(z_0)}e^{6A(z)-6A(z_0)}}}-1\Bigg)} \Biggr]}
\end{equation}
\begin{equation}\label{eq88}
\frac{\hat{U}^2_2}{\hat{U}_1^2} =\exp{\Bigg[\frac{4\pi}{\beta}\int_{0}^{z_H}dz~{\frac{R^2 e^{B(z)-A(z)}}{z^2 h(z)}\Bigg(\frac{1}{\sqrt{1-\frac{h(z)}{h(z_0)}e^{6A(z)-6A(z_0)}}}-1\Bigg)} \Bigg]}
\end{equation}

\begin{figure}
\centering
 \includegraphics[width=.50\linewidth]{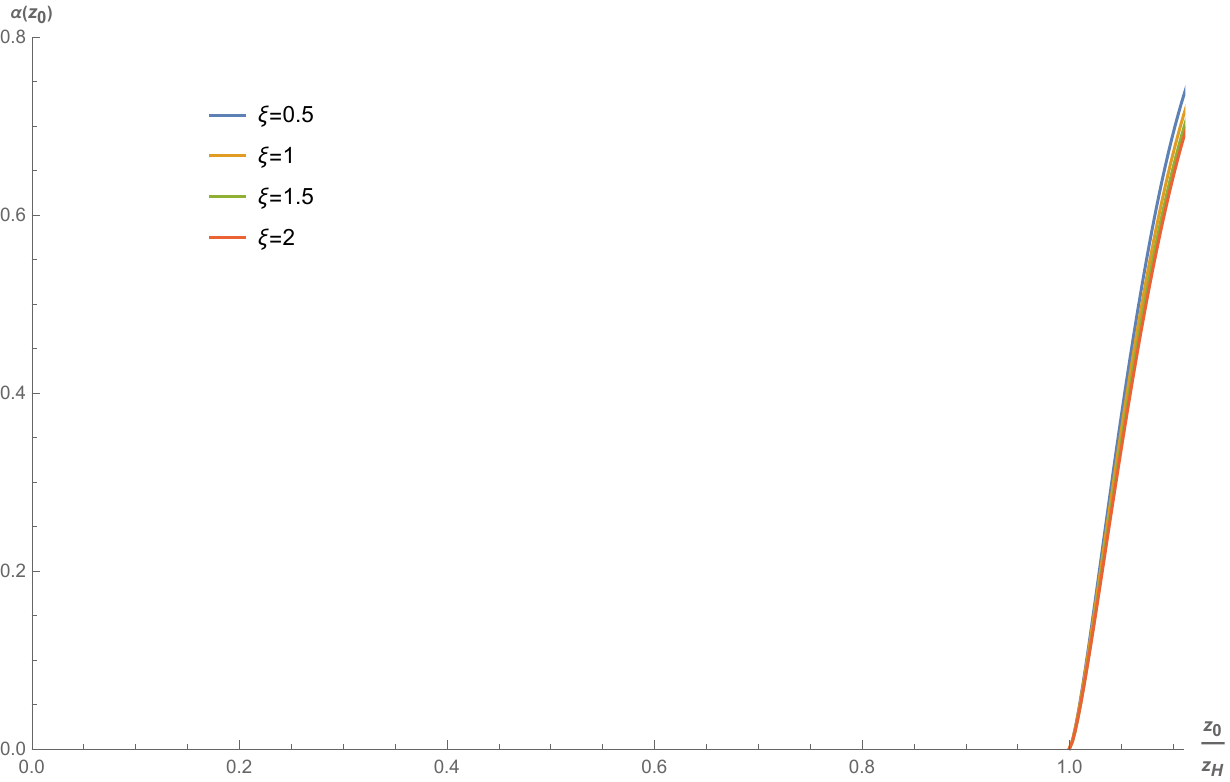}
 \caption{Shockwave parameter $\alpha (z_0)$ v/s $\frac{z_0}{z_H}$ for different values of $\xi$.}
\label{alphaz0}
\end{figure}

To find $\hat{U}_2$, we consider a reference point at $\Bar{z}$ where $z_*$ is zero.
\begin{equation}\label{eq89}
\hat{V}_2\hat{U}_2 =\exp{\left(\frac{4\pi}{\beta}z_*\right)}=\exp{\Bigg[-\frac{4\pi}{\beta}\int_{\bar{z}}^{z_0} dz~{\frac{R^2e^{B(z)-A(z)}}{z^2h(z)}}\Bigg]}
\end{equation}
In the region (III), where $\dot{z}>0$, the $h(z)$ remains in the negative numerical range, we introduce an overall negative sign to the variable $t$. As a result, the expression for the coordinate $\Delta \hat{V}$ in region (III) adopts the following form:
\begin{equation}\label{eq90}
\frac{\alpha^2}{4\hat{V}_2^2} =\frac{4\pi}{\beta}\int_{0}^{z_h} dz~{\frac{R^2 e^{B(z)-A(z)}}{z^2 h(z)}\Bigg(\frac{1}{\sqrt{1-\frac{h(z)}{h(z_0)}e^{6A(z)-6A(z_0)}}}-1\Bigg)} 
\end{equation}
By combining equation \eqref{eq89} and \eqref{eq90} we establish a relation between $\alpha$ and $z_0$ as,
\begin{equation}\label{eq91}
\alpha(z_0) =2\exp{(\mathcal{\eta}_{\text{I}}+\mathcal{\eta}_{\text{II}}+\mathcal{\eta}_{\text{III}})} 
\end{equation}
where
\begin{equation}\label{eq92}
\begin{split}
\mathcal{\eta}_{\text{I}} &=\frac{4\pi}{\beta}\int_{\bar{z}}^{z_0} dz~ {\frac{R^2e^{B(z)-A(z)}}{z^2h(z)}}
,~~~~\mathcal{\eta}_{\text{II}}=\frac{2\pi}{\beta}\int_{0}^{z_h} dz~ {\frac{R^2 e^{B(z)-A(z)}}{z^2 h(z)}\Bigg(\frac{1}{\sqrt{1-\frac{h(z)}{h(z_0)}e^{6A(z)-6A(z_0)}}}-1\Bigg)} \\
\mathcal{\eta}_{\text{III}}&=\frac{4\pi}{\beta}\int_{z_h}^{z_0} dz~ {\frac{R^2 e^{B(z)-A(z)}}{z^2 h(z)}\Bigg(\frac{1}{\sqrt{1-\frac{h(z)}{h(z_0)}e^{6A(z)-6A(z_0)}}}-1\Bigg)} 
\end{split}
\end{equation}

\begin{figure}
\centering
\includegraphics[width=.50\linewidth]{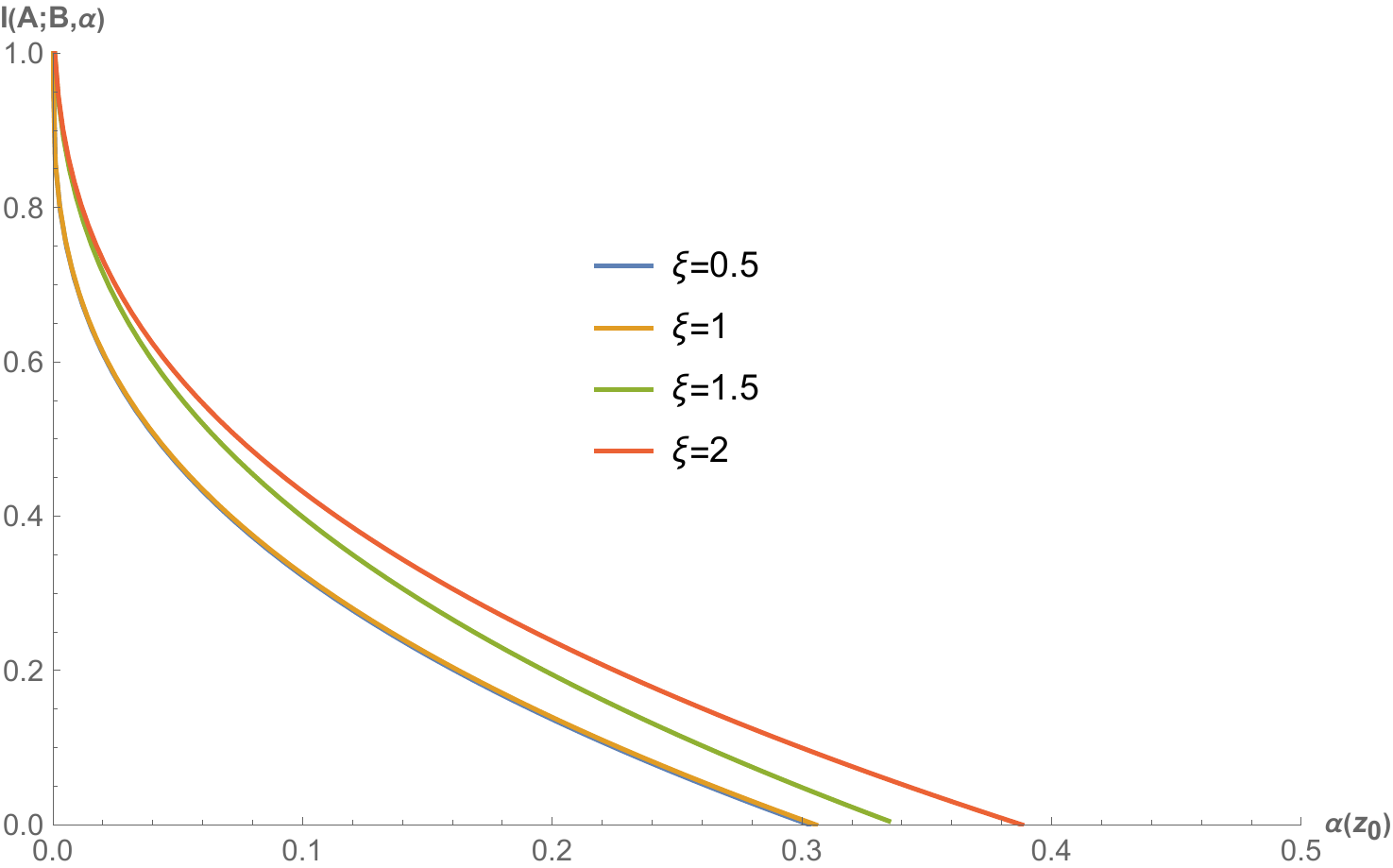}
\caption{TMI v/s shock parameter $\alpha$ for different $\xi$ and $T=1, R=1$.}
\label{MIvsalpha}
\end{figure}

From equation \eqref{eq91}, it becomes possible to generate a graphical representation of the shock wave parameter concerning the dimensionless quantity $z_0/z_H$, as depicted in Fig.\ref{alphaz0}. This is in accordance with the expectation that the shockwave parameter $\alpha$ increases as the parameter $z_0$ increases. This increment in $\alpha$ with respect to the $z_0$ depends on $\xi$. For instance, when $\xi$ takes larger values, the rate at which $\alpha$ increases is comparatively slower in contrast to situations for smaller $\xi$ values.
In the end, in Fig. \ref{MIvsalpha}, the graph illustrates how the HTMI changes concerning the shockwave parameter $\alpha$ for various nonzero $\xi$ values. The HTMI starts decreasing from a specific value as $\alpha$ increases. The rate at which the HTMI decreases is controlled by the parameter $\xi$. For small non zero values of $\xi$, the HTMI decreases more rapidly rather then the larger values of $\xi$ as depicted from \ref{MIvsalpha}, the HTMI only exists for $\alpha$ values less than or equal to $\alpha_c$. This critical value of $\alpha$ increases as the $\xi$ parameter grows. Therefore it's evident that HTMI remains finite at the critical point.  A similar observation is made for one-sided HMI in \cite{Ebrahim:2020qif}.

\section{Summary and Discussions}\label{sec:sec9}

In this work, we study the various measures for the entanglement structure of mixed states and the properties of chaos in the four-dimensional $\mathcal{N} = 4$ super Yang-Mills theory at finite temperature T, charged under a $U(1)$ subgroup of its $SU(4)$ R-symmetry. In particular, we analytically explore the HLN, EWCS in the low and high temperature limit to probe the entanglement structure near the critical point $\xi \to 2$. We  analyse the two sided mutual information HTMI and the disruption of HTMI due the shockwave perturbation and finally interpret our results in terms of boundary theory parameters.

We study the influence of the dimensionless parameter $\xi$ related to the charge of black hole on the HLN in low and high temperature limits. In this analysis, we observe that the corresponding RT surface dual to the boundary region A receives a modification due to the presence of $\xi$. Moreover, for a fixed width $l$ of the boundary region $A$, the RT surface goes deeper into the bulk for larger value of $\xi$.
For computing the HLN at low and high temperature, we consider adjacent, disjoint and bipartite configuration of subsystems in the boundary. As a consistency check we take a limit $\xi \to 0$ ($Q\to 0$) on our analysis and correctly reproduce the results obtained for the AdS$_{d+1}$ Schwarzschild black hole background. The HLN exhibits an increasing behavior for adjacent configurations with respect to the parameter $\xi$ both at low and high-temperature regimes. For disjoint subsystems, the HLN increases with $\xi$ for low temperature and vanishes for high temperature. In our analysis, bipartite subsystems follow the same behavior exhibited by the disjoint subsystems  at low temperature.
In the field theory, the growth of the HLN can be understood as indicative of the increasing entanglement between two subsystems.
As the critical limit is approached ($\xi \to 2$) in all cases, HLN remains finite while the slope of the HLN shows a power law divergence at the critical point $\lambda\to 1$. A similar finding was previously documented for HEE and HMI in \cite{Ebrahim:2020qif}.

We give analytic expressions for EWCS for 1RC black hole in the low and high temperature limits. Our result turns out to be consistent with the numerical result previously obtained in \cite{Amrahi:2020jqg}. 
We observe that, at low temperatures, the EWCS experiences a correction attributed to the parameter $\xi$ and consequently it exhibits a growth with respect to $\xi$. We observed that EWCS is finite in the critical limit and the slope of EWCS shows a power law divergence. The relation between the mutual information and the EWCS discussed in \cite{Umemoto:2019jlz} is useful to show the agreement of our EWCS result with the HMI result reported in \cite{Ebrahim:2020qif} for the case of 1RC black hole.
For disjoint case, in the low-temperature regime, both the HLN and EWCS exhibit a similar dependence on the boundary parameter (characteristic length of the different regions) as well as the temperature. In the high-temperature limit, these quantities vanish as stated in \cite{Ebrahim:2020qif}.

We notice that the entanglement between two subsystems of a TFD state measured by TMI increases with the size of the subsystem. If we fix the size of subsystems, TMI increases as  $\xi$ parameter approaches to the higher values. Based on our analysis, we conclude that the two separate subsystems do not manifest correlations regardless of the subsystem's size. It is only once a specific size, denoted as the critical width $l_c$, is reached then the total correlation starts to emerge. 
We demonstrate the explicit disruption of holographic TMI in the presence of a shockwave. Our finding suggests that the parameter $\xi$ attempts to slow down the disruption, indicating that the presence of the $\xi$ parameter tends to reduces the chaotic behavior of the system.
For a substantial value of $\xi$, holographic TMI exhibits a slower rate of decay. In other words, when $\xi$ is large, it takes more time for TMI to completely vanish.
 
 In summary, we show that the parameter $\xi$ enhances the correlation in the strongly coupled field theory of our interest and at the critical point $\xi \to 2$ all the correlation measures we compute are finite and well behaved. We argue that similar to the mutual information,  HLN and EWCS could also be used in order to study the critical phenomena of the strongly coupled field theories in the large-N limit due to their power law behaviour near the critical point. On the other hand by increasing the value of $\xi$ one can slow down the rate of disruption of TMI.

\bigskip
\centerline{\bf Acknowledgement}
We express our gratitude to Shankhadeep Chakrabortty for valuable comments on the draft. SP acknowledges the support of Senior Research Fellowship from the Ministry of Human Resource and Development, Government of India. SP expresses gratitude to Dongmin Gang and Seoul National University for their generous support and warm hospitality during a part of this work. DK wishes to extend appreciation to Shankhadeep Chakrabortty and IIT Ropar for their support and warm hospitality during the final stages of this project.

\appendix
\section{Appendix:}
\subsection{Area of the Extremal Surface for Bipartite Systems}\label{app:A}
We provide a concise overview of employing the near horizon expansion technique to estimate extremal surfaces within the bipartite subsystem, aiming to determine the surface areas $\mathcal{A}_{B_1}$ and $\mathcal{A}_{A\cup B_1}$ in the limit $L \to \infty$. It is convenient to start with the equation \eqref{eq25} by rewriting it in the following form
\begin{equation}\label{eqA1}
    \mathcal{A} = {\mathcal{A}}^{(1)} + {\mathcal{A}}^{(2)} + {\mathcal{A}}^{(3)}
\end{equation}
where we define the quantities ${\mathcal{A}}^{(1)}, {\mathcal{A}}^{(2)}, {\mathcal{A}}^{(3)}$ in the following way
\begin{equation}\label{eqA2}
    {\mathcal{A}}^{(1)} = \frac{L^2 R^3}{{z_t}^2}\Bigg{\{} \frac{3\xi}{2} {\left( \frac{z_t}{z_h}\right)}^2 - {\left[ 1 + \xi {\left( \frac{z_t}{z_h}\right)}^2\right]}^{\frac{3}{2}} + \frac{1+\xi}{3\xi}{\left( \frac{z_t}{z_h}\right)}^2 \left[ {\left( 1 + \xi {\left( \frac{z_t}{z_h}\right)}^2 \right)}^{\frac{3}{2}} - 1 \right]\Bigg{\}}
\end{equation}
\begin{equation}\label{eqA3}
    {\mathcal{A}}^{(2)} = \frac{L^2 R^3}{{z_t}^2} \Bigg{\{} \sum_{n=0}^2  {\Lambda}_{2n0} \frac{ \sqrt{\pi}\Gamma(n+1)}{\Gamma(n+3)} {\left( \frac{z_t}{z_h} \right)}^{4+2n}  \\ \times \left[ 1 + (n+1) \left( 1 + \xi{\left( \frac{z_t}{z_h}\right)}^2 \right) \right]  \Bigg{\}}
\end{equation}
\begin{equation}\label{eqA4}
    {\mathcal{A}}^{(3)} = \frac{L^2 R^3}{{z_t}^2} \Bigg{\{}  \sum_{j=1}^{\infty} {\Lambda}_{000} \frac{ \Gamma(j+\frac{1}{2})\Gamma(3j-1)}{\Gamma(j+1)\Gamma(3j+1)} \times \left[ 1 + (3j-1) \left( 1 + \xi{\left( \frac{z_t}{z_h}\right)}^2 \right) \right]  \Bigg{\}}
\end{equation}
Note that we have truncated the series present in ${\mathcal{A}}_2$ and ${\mathcal{A}}_3$ in order to obtain the lowest-order contribution, signifying the near-horizon expansion. 
We then employ equation \eqref{eq61} within the context of equations \eqref{eqA2}, \eqref{eqA3}, and \eqref{eqA4} to derive the expression for extremal surfaces in the bipartite limit, as presented in equation \eqref{eq64}. In this context, we introduce the following functions of $\xi$.

\begin{equation}\label{eqA5}
    \alpha(\xi) = \left( \frac{3\xi}{2}-\frac{2}{3}(1+\xi)+\frac{k_2}{3\xi}(1-2\xi)(\xi-2) \right),~~ \beta(\xi)=k_2\sqrt{6}\frac{(2\xi -1)(\xi -2)}{3\xi}
\end{equation}
\begin{equation}\label{eqA6}
\begin{split}
&\gamma(\xi) = \Bigg( u_1(\xi)-v_1(\xi)+x_1(\xi)+k_2\bigg(u_2(\xi)-v_2(\xi)+x_2(\xi) \bigg)\Bigg),~~ \delta(\xi)= \Bigg(-u_2(\xi)+v_2(\xi)-x_2(\xi)\Bigg) \\ &u_1(\xi) = \frac{3\xi^2}{16}(\xi^2 + 3\xi +2) \hspace{2 cm} u_2(\xi) = \frac{3\xi^2}{16}(3\xi^2+6\xi+4) \\ &v_1(\xi) = \frac{3\xi(1+\xi)}{24}(2\xi^2 + 5\xi +3) \hspace{7 mm} v_2(\xi)=\frac{3\xi(1+\xi)}{24} (10\xi^2 + 21\xi +12)\\ &x_1(\xi) = \frac{3{(1+\xi)}^2}{96}(\xi^2 +7\xi+4) \hspace{8 mm}  x_2(\xi) = \frac{3{(1+\xi)}^2}{96} (21\xi^2 + 44\xi + 24)
\end{split}
\end{equation}
\begin{equation}\label{eqA7}
\begin{split}
    \mu(\xi)= \sum_{j=1}^\infty \frac{3}{\sqrt{\pi}}\frac{\Gamma\left( j+\frac{1}{2}\right)\Gamma(3j-1)}{\Gamma(j+1)\Gamma(3j+1)}(1+\xi-(\xi+2)k_2) \\
    \nu(\xi)= \sum_{j=1}^\infty \frac{3}{\sqrt{\pi}}\frac{\Gamma\left( j+\frac{1}{2}\right)\Gamma(3j-1)j}{\Gamma(j+1)\Gamma(3j+1)}(\xi +2)k_2\sqrt{6}
\end{split}
\end{equation}
Using the above functions we obtain the following extremal areas
\begin{equation}\label{eqA8}
\begin{split}
    {\mathcal{A}}_{B_1} = \frac{L^2R^3}{z_h^2}\Big{\{} \alpha(\xi)+\gamma(\xi)+\mu(\xi) \Big{\}} + \frac{L^2R^3}{z_h}\left( L-\frac{l}{2}\right) \Big{\{}  \beta(\xi) +\delta(\xi) + \nu(\xi) \Big{\}} \\
     \mathcal{A}_{A\cup B_1} = \frac{L^2R^3}{z_h^2}\Big{\{} \alpha(\xi)+\gamma(\xi)+\mu(\xi) \Big{\}} + \frac{L^2R^3}{z_h}\left( L+\frac{l}{2}\right) \Big{\{}  \beta(\xi) +\delta(\xi) + \nu(\xi) \Big{\}}
\end{split}
\end{equation}
By applying the equations mentioned earlier to the entanglement negativity formula associated with the bipartite state in a low-temperature regime, we derive equation \eqref{eq65}. It is worth noting that \eqref{eq65} is obtained through the relation between $\hat{T}$ and $z_h$. Additionally, we introduce the following function for a more concise expression of the HLN.

\begin{equation}\label{eqA9}
g(\xi) = \pi \big( \beta(\xi) +\delta(\xi) + \nu(\xi)\big)
\end{equation}
We employ the defined function $g(\xi)$ to examine the scenario as we approach the limit where $\xi$ tends towards zero. In this limit, it becomes evident that \eqref{eqA9} simplifies to the following expression.
\begin{equation}\label{eqA10}
    g(\xi){\Bigg{\vert}}_{\xi\to 0} = \pi \Bigg( \frac{2}{3}k_2\sqrt{6}\frac{1}{\xi} - \frac{3}{4} + 2k_2\sqrt{6}\sum_{j=1}^\infty \frac{3}{\sqrt{\pi}}\frac{\Gamma\left( j+\frac{1}{2}\right)\Gamma(3j-1)j}{\Gamma(j+1)\Gamma(3j+1)} \Bigg)
\end{equation}

 The temperature dependence of the function $g(\xi)$ in the limit $\xi \to 0$ can be considered by the first term inside the parenthesis in the above equation and this shows that the function $g(\xi)$ is proportional to $T^2$. The constant $\mathcal{C}$ in \eqref{eq66} can be obtained by taking the coefficient of the first term in equation \eqref{eqA10} into account

\subsection{Approximate EWCS at low temperature limit in terms of boundary parameters}\label{app:B}
In this appendix, we derive the expression for the EWCS under the conditions of low temperature. By inserting the equations for the turning points into the \eqref{ewcs5} the expression for the EWCS in the series form is,

\begin{equation}
   \begin{split}
       E_W^{low} = \frac{L^2R^3}{4G_N^5} \sum_{k=0}^{\infty} \sum_{j=0}^{k} \sum_{i=0}^{\infty} \frac{{(-1)}^{k+j}}{2}\frac{\Gamma(k+\frac{1}{2}) {\xi}^{i+j+k}{(1+\xi)}^j}{\Gamma(i+1)\Gamma(\frac{3}{2}-i)\Gamma(j+1)\Gamma(k-j+1)}\frac{z_h^{2-2i-2j-2k}}{(2i+2j+2k-2)} \\
        \times \Bigg[ \Bigg{\{} {\left( \frac{2l+D}{a_1}\right)}^{2i+2j+2k-2} - {\left(\frac{D}{a_1}\right)}^{2i+2j+2k-2}\Bigg{\}} + (2i+2j+2k-2)\frac{\xi}{6z_h^2}\Bigg{\{}{\left(\frac{2l+D}{a_1}\right)}^{2i+2j+2k} \\
        - {\left(\frac{D}{a_1}\right)}^{2i+2j+2k}\Bigg{\}} + \frac{2i+2j+2k-2}{2z_h^4}\left(\frac{{\xi}^2}{6}\left( 1-\frac{a_3}{2a_2} \right) \right) \Bigg{\{} {\left(\frac{2l+D}{a_1}\right)}^{2i+2j+2k+2} - {\left(\frac{D}{a_1}\right)}^{2i+2j+2k+2} \Bigg{\}} \\ 
        + \mathcal{O}{\left(\frac{1}{z_h} \right)}^6
        \Bigg]
    \end{split}
\end{equation}

As stated in Section \ref{subsec:sec8.1}, we have the option to conclude the series for increased simplification by setting $i=j=k=0$. This process simply gives the expression \ref{ewcs9}. 

\end{document}